\documentclass[preprint,12pt,short&compress]{elsarticle}
\usepackage{amssymb,amsmath}
\usepackage{multirow}
\usepackage{array}
\usepackage{diagbox}
\usepackage{multirow}
\usepackage{xcolor}
\usepackage{float}
\begin{document}

\begin{frontmatter}



%\title{Role of periodic and wall boundary conditions on dam-break flow across an obstacle}
\title{Role of boundary conditions on dam-break flow across an obstacle and controlling damage of structures}
% \author{Hari}
 %\affiliation{organization={iiit}},
%\author[mymainaddress]{P. C. Harisankar}
%\author[mymainaddress]{{Tapas Sil}\corref{mycorrespondingauthor}}%
%\cortext[Tapas Sil]{Corresponding author}
%\ead{tapassil@iiitdm.ac.in}
\author[mymainaddress]{P. C. Harisankar}
\author[mymainaddress]{{Tapas Sil}\corref{mycorrespondingauthor}}%
\cortext[Tapas Sil]{Corresponding author}
\ead{tapassil@iiitdm.ac.in}
\affiliation{organization={Department of Physics, Indian Institute of Information Technology, Design and Manufacturing Kancheepuram},%Department and Organization
	addressline={Chennai-600127, Tamil Nadu, India}}
%\address[mymainaddress]{Department of Physics, Indian Institute of Information Technology, Design and Manufacturing Kancheepuram, Chennai-600127, Tamil Nadu, India }
%\affiliation{organization={},%Department and Organization
\begin{abstract}
We studied dam-break flow in the smoothed particle hydrodynamics framework using periodic boundary condition (PBC) instead of usually employed rigid wall boundary condition (WBC) and assessed the effects of impact of the flow on the downstream structure due to the presence of an obstacle in front of it.  The results show that higher dam heights lead to larger pressure on the wall. The WBC yields higher peak pressures compared PBC. A larger hydraulic diameter of the pillar is found to be more efficient in reducing the flow's impact. A pillar located closer to the wall reduces the effect of dam-break flow and minimises structural damage. The square-shaped pillars are found to be the most effective in reducing pressure on the wall among the considered pillar shapes. These findings will help to mitigate the damage of a structure due to dam-break flow/high-tide and improve the safety of the structures downstream.
These findings have direct implications for the design and management of structures in areas prone to dam-break flows. 
%
%\begin{description}
	%\item[PACS]%47.11.+j, 47.55.Dz, 02.60.-x, 02.70.Ns$
%\end{description}
\end{abstract}
\begin{keyword}
Dam-break \sep Smoothed Particle Hydrodynamics (SPH) \sep Periodic Boundary Condition (PBC) \sep Wall Boundary Condition (WBC) \sep Obstacle \sep Impact
\end{keyword}

\end{frontmatter}


\section{Introduction}
The dam-break flow serves as a benchmark problem \cite{zhou1999,hu2004cip} in fluid dynamics simulations, where a liquid column undergoes free fall under gravity, replicating the flow resulting from the sudden release of water from a breached dam. Such a setup is also widely used as a wave-maker to analyse the impact of water flow during natural disasters, such as tsunamis and floods, on surrounding structures and terrain. There are two aspects of such studies: flow dynamics and fluid-structure interaction.
Several studies have been reported on dam-break flow in the context of fluid-structure interaction \cite{marrone2011,hien2021,begnudelli2007,aureli2015,elkholy2016exp,soares2007exp,hu2010exp,kamra2019exp}, which provide valuable insights into the behaviour of fluid forces and structural responses under sudden flow conditions. 

Numerical simulations of fluid dynamics have been studied using various methods \cite{Maranzoni2023}. The Finite Difference Method (FDM) \cite{narasimhan1976} approximates derivatives by using differences between function values at discrete grid points, which is straightforward and easy, but does not guarantee conservation of mass, momentum, and energy on coarse grids, and is not efficient for modelling complex geometries.
The Finite Volume Method  (FVM) \cite{barth2018finite,vieira2024face} involves dividing the fluid domain into small control volumes and applying the conservation laws to each volume. This method is notably better for handling complex geometries and boundary conditions. Another method for such simulation is the  Finite Element Method (FEM) \cite{ZHANG2018}, where the computational domain is discretised into elements and the solution is obtained by minimising the error function using variational methods. FEM is versatile and efficient for specific problems, such as simulating flows with complex boundary conditions and multi-phase flows. But it is often computationally more intensive. Direct Numerical Simulation (DNS) solves the governing equations of fluid flow directly; however, turbulence is not considered \cite{scardovelli1999}. It is highly accurate but computationally expensive, making it suitable mainly for simple flows or small-scale simulations. Large Eddy Simulation (LES) is a computationally more efficient method for simulating turbulent flows \cite{hien2021}. It resolves the larger scales of turbulence while modelling the smaller scales, providing a balance between accuracy and computation time. The Reynolds-Averaged Navier-Stokes (RANS) equations method involves averaging the Navier-Stokes equations to simplify the analysis of turbulent flows. RANS models are commonly used for industrial applications where turbulence plays a significant role \cite{xiong2016rans}.

The Smoothed Particle Hydrodynamics (SPH) method is a mesh-free simulation technique that was developed in 1977 \cite{lucy1977,monaghan1985} for solving problems associated with astrophysical phenomena. SPH was later adapted for various branches of science and engineering, such as solid mechanics \cite{avs2015numerical}, fluid dynamics \cite{adami2012}, liquid drop and vapour system \cite{harisankar2023drop,nugent2000,ray2017,yang2014} and heat transfer problems \cite{liu2005,nasiri2019,harisankar2025smoothed,sil2023nanofluid}. Being a particle-based meshfree method, it offers advantages in handling free-surface flows, large deformations, and complex boundary interactions, which are modelled easily and efficiently \cite{liu2003,harisankar2023drop}. 
The existing numerical studies primarily analyse the direct impact of dam-break flows on objects in the downstream.

%Kedarnath, one of the most sacred temples of India, was saved by a stone!
During the 2013 flash flood, the mountain behind the Kedarnath temple (Uttarakhand, India) broke away, sending rampaging waters through the Kedarnath valley. Destroying everything in its path, the Kedarnath temple was spared as one gigantic stone (about 20 feet wide and 12 feet tall) rolled down the mountain and stopped a few meters behind the temple, causing the water flow to split and rush around the sides of the temple and flash over the temple.
It will be interesting to investigate systematically the impact of the dam-break flow on the object in the downstream, considering the effects of an obstacle in front of it. Usually, the impact directly on downstream objects is reported in the literature \cite{Watanabe2021,mu2023,aureli2015}. 
We investigate how an obstacle alters flow behaviour and evaluate the resulting forces and pressure distributions on the wall located behind it. 

 The periodic boundary conditions (PBC) is applied to a system where a periodicity exists \cite{ellero2011sph,wang2019analytical,Mardanov2016Periodic} and are profusely used in simulating a continuous flow, where the periodicity is along the direction of flow \cite{Kim2007GeneralizedPBC,Jonsson2016SPH}.  In case of a wide flow as seen in sea, river and wide dam-break flow, we see the sidewise periodicity (transverse to the flow direction) if we consider a  portion (workspace) of width. Usually imaginary side walls considered with no slip condition to model such problems \cite{Agamloh2008FSIWaveEnergy,moore2004lateral,harisankar2025smoothed,sil2023nanofluid}. But practically, fluid particles may come in and go out of the workspace through the imaginary side-walls. Therefore, we consider PBC to model such systems. We employ the SPH method to model dam-break flow with periodic boundary conditions (PBC) incorporating variations of the parameters (size, location, and shapes) of the obstacle (pillar) to analyse their influence on the impacting forces on the wall. The results obtained from PBC  are compared with those from rigid wall boundary conditions plus slip condition 
(WBC) to evaluate the effect of these boundary conditions on the dynamics of the fluid and the forces exerted on the downstream wall. This comparison provides insights into how boundary conditions influence flow dynamics, pressure distribution, and energy dissipation within the system.

A brief of SPH formalism is given in section 2. In Section 3, we discuss the results obtained from our calculations for various dam-break conditions. Results are summarised and concluded in section 4.
\section{Formulation}
\subsection{Smoothed particle hydrodynamics}
Smoothed particle hydrodynamics is developed based on the Lagrangian method, which tracks individual fluid particles as they move through space and time, making it particularly well-suited for capturing free surface flows, interface dynamics, and particle-based phenomena. The SPH method consists of two approximations: kernel approximation and particle approximation.

In the Kernel approximation, a kernel function is considered, which has a delta function-like behaviour. Any given function $f(\mathbf{x})$ and its derivatives are expressed in continuous form as an integral representation. The kernel approximation of a function and its derivatives is \cite{liu2003},
\begin{equation}
	\langle f(\mathbf{x})\rangle = \int_\Omega f(\mathbf{x}^\prime)W(\mathbf{x} - \mathbf{x}^\prime,h)\mathbf{dx}^\prime,
	\label{eq:funAvg}
\end{equation} 
\begin{equation}
	\langle\nabla f(\mathbf{x})\rangle = -\int_\Omega f(\mathbf{x}^\prime)\nabla W(\mathbf{x} - \mathbf{x}^\prime)\mathbf{dx}^\prime,
	\label{eq:GradfunAvg}
\end{equation} 
where $h$ is the smoothing length, which is ideally taken equal to the initial particle spacing $dx$, $ W(\mathbf{x} - \mathbf{x}^\prime,h)$ is the kernel function, and is defined considering neighbouring particles within a distance $kh$, $k$ being an integer. 
The particle approximation represents the problem domain using a set of particles and estimates observables, such as velocity, density, and stress, for this set of particles. 
Then the particle approximation for a function $f(\mathbf{x})$ and its gradient for the particle at  $a$ are discretised and expressed as a weighted summation over its neighbours and are given by,
\begin{equation}
	\langle f(\mathbf{x}_a)\rangle =\sum_b \frac{m_b }{\rho_b}f(\mathbf{x}_b)W_{ab}
	\label{eq:funSPH},
\end{equation}
\begin{equation}
	\langle \nabla f(\mathbf{x}_a)\rangle =\frac{1}{\rho_a}\left[\sum_b m_b(f(\mathbf{x}_b)-f(\mathbf{x}_a))\nabla W_{ab}\right],
	\label{eq:funAvgSPH}
\end{equation}
where the suffix $a$ and $b$ represent two SPH particles. In Eq. (\ref{eq:funSPH}), $\rho$ represents density, $m$ is the mass of the SPH particle. 
We consider the B-spline kernel function based on the cubic spline \cite{monaghan1985}, which is written as,
\begin{equation}
	W_{ab}=\alpha_d% 
	\begin{cases}
		\frac{2}{3}-s^2+\frac{1}{2}s^3, & 0\le s <1\\
		\frac{1}{6}(2-s)^3, & 1\le s <2\\
		0, &  s \ge 2,\\
	\end{cases}
	\label{eq:Kernal}
\end{equation}
where for 3D,
\begin{equation}
	\alpha_d=\frac{3}{2\pi h^3} \:\: \text{and} \:\: 	s=\frac{|\mathbf{x}_a-\mathbf{x}_b|}{h}. 
	%\begin{cases}
	%\frac{1}{h}, & for\: 1\, D,\\
	%\frac{15}{7\pi h^2}, & for\: 2\,  D,\\
	%\frac{3}{2\pi h^3}, & for\: 3\,  D.
	%\end{cases},\:\: and \:\: 	%s=\frac{|\bold{r}_a-\bold{r}_b|}{h},
	\label{eq:Kerneldim}
\end{equation}
Here, 	$\mathbf{x}_a$ and $\mathbf{x}_b$ are the position vectors of particle $a$ and particle $b$. 
The Navier-Stokes equation for conservation of mass and momentum  is given as, \cite{adami2012}.
\begin{equation}\label{eq:continuity}
	\frac{D\mathbf{\rho}}{Dt}=-\rho\nabla.\mathbf{u},
\end{equation}
\begin{equation}\label{eq:momentum}
	\frac{D\mathbf{u}}{Dt}=-\frac{1}{\rho}\mathbf{{\nabla}}P+\frac{\mu}{\rho}\nabla^2 \mathbf{u}+\mathbf{g},
\end{equation}
where, $\mathbf{u}$ is the velocity, $P$ is pressure, $\mu$ is viscosity, $\mathbf{g}=(0,0,-9.8m/s)$ is acceleration  due to the body force $\mathbf{F_b}$ (here gravity). The pressure is calculated using Tait's equation of state,
\begin{equation}
	P = \frac{c_{0}^2\rho_0}{\gamma}\left\lbrace
	\left(\frac{\rho}{\rho_0}\right)^{\gamma} -1\right\rbrace,
	\label{eq:TaitEOS}
\end{equation}
where $c_0=10u_{max}$, ($u_{max}=\sqrt{2gH}$, is the maximum fluid velocity,)  is the speed of sound, $\rho_0$ is the reference density, and $\gamma$ is a polytropic constant, which varies from 1 to 7, for an example, for water $\gamma=7$.
%where $p_b=\frac{c_{0}^2\rho_0}{\gamma}$.
The SPH discretised form of Eq.(\ref{eq:continuity}), and Eq.(\ref{eq:momentum}) is given as, \cite{hu2006multi,adami2012}, 
\begin{equation}
	\frac{D\mathbf{\rho_{a}}}{Dt}=\rho_a\sum_b \frac{m_b}{\rho_b}(\mathbf{u}_a-\mathbf{u}_b).\mathbf{\nabla W_{ab}},
	\label{eq:ContSPH}
	\underline{}
\end{equation}
\begin{eqnarray}
	\begin{split}
		\frac{D\mathbf{u}_{a}}{Dt}=&-\frac{1}{m_a}\sum_b \left(V_{a}^{2}+V_{b}^{2}\right)\tilde P_{ab}\mathbf{\nabla W_{ab}}\\
		&+\frac{1}{m_a}\sum_b \left(V_{a}^{2}+V_{b}^{2}\right)\tilde \mu_{ab}\frac{\mathbf{\mathbf{x}}_{ab}.\mathbf{\nabla W_{ab}}}{|\mathbf{x}_{ab}|^{2}+\epsilon h^2}\mathbf{u}_{ab}
		+\mathbf{g},\\
		\label{eq:MomSPH}
	\end{split}
\end{eqnarray}
\begin{figure}[H]
\centering
\includegraphics[width=1\textwidth]{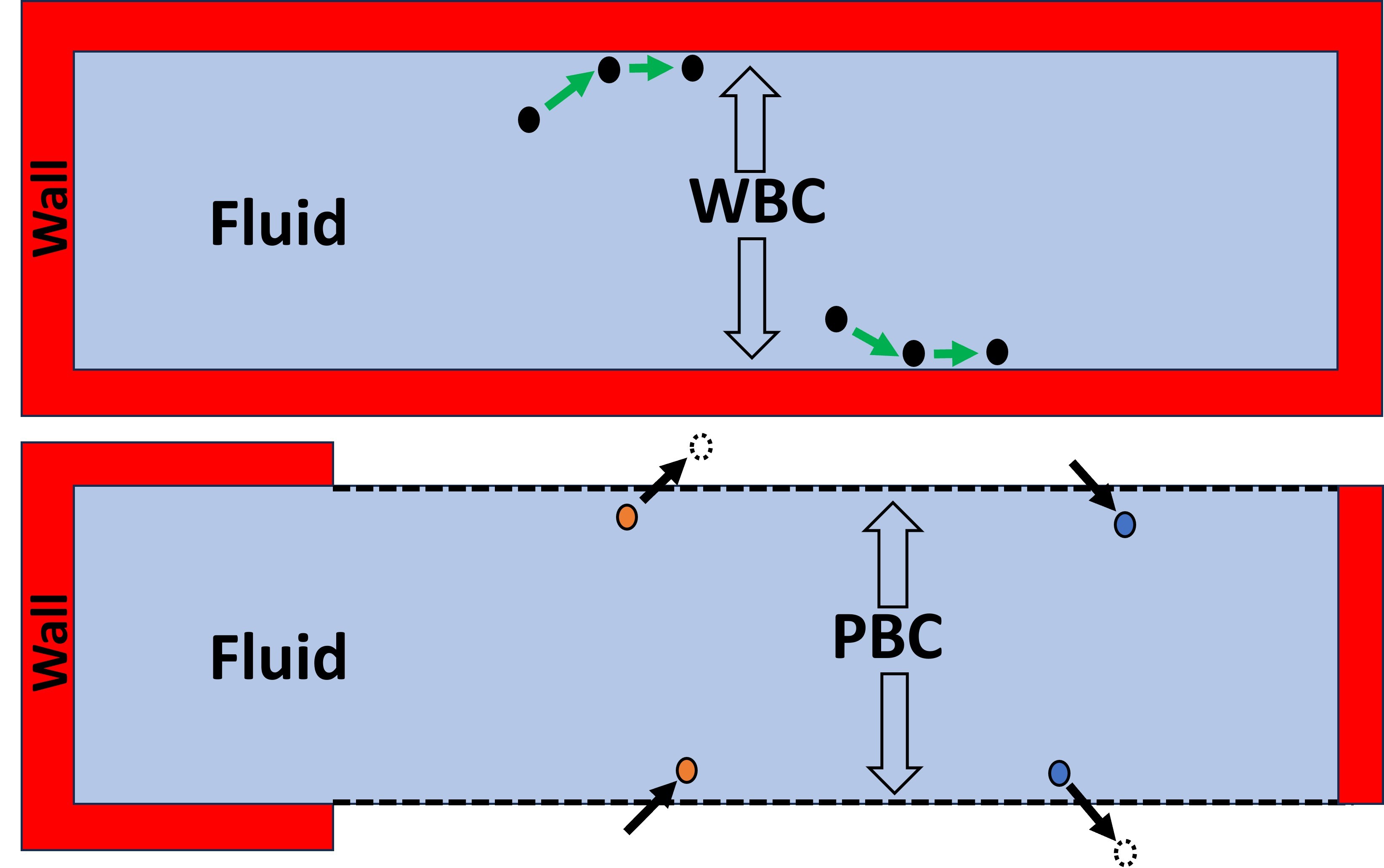}\\
	\caption{Schematic diagram of periodic boundary condition (PBC) and Wall boundary condition (WBC).}
	\label{Fig:PBC}
\end{figure}
where, $\tilde P_{ab}=\frac{P_a\rho_a+P_b\rho_b}{\rho_a+\rho_b}$; $\tilde \mu_{ab}=\frac{\mu_{a}\mu_{b}}{\mu_{a}+\mu_{b}}$; $V_i=\frac{m_i}{\rho_i}$, is volume of the particle; $\mathbf{x}_{ab}=\mathbf{x}_{a}-\mathbf{x}_{b}$ and  $\mathbf{u}_{ab}= \mathbf{u}_a-\mathbf{u}_b$.

The XSPH is adapted which  updates velocity field as \cite{monaghan1989xsph}, 
\begin{eqnarray}
	\frac{d\mathbf{r}_{a}}{dt}=\mathbf{u}_{a}+
	\epsilon\sum_{b}m_{b}\frac{\mathbf{u}_{ab}}{\bar{\rho}_{ab}}W_{ab}.
	\label{eq:XSPH}
\end{eqnarray}
This correction helps to get a stable fluid flow.
The fluid-solid interaction is considered \cite{adami2012,song2018} by iterating the fluid velocity and pressure to the dummy wall particles,
\begin{equation}
	\mathbf{u}=2\mathbf{u}_w-\hat{\mathbf{u}},
\end{equation}
where $\mathbf{u}_w$ is the real wall velocity and $\hat{\mathbf{u}}$ is,
\begin{equation}
	\hat{\mathbf{u}}=\frac{\sum \mathbf{u}_{f}W_{f}}{\sum W_{f}},
\end{equation}
Similarly wall pressure is calculated as,
\begin{equation}
	P=\frac{\sum P_{f}W_{f}}{\sum W_{f}}.
\end{equation}

The properties of SPH particles are updated at each time step $dt$  ($n$ is the number of steps).
%
% using the following equation:
%\begin{eqnarray}\label{update}
%	%\begin{array}
%	\bold{u}_a^{n+1}=\bold{u}_a^{n}+\frac{d\bold{u}}{dt}^{n+1}\times dt,\\
%	\rho_a^{n+1}=\rho_a^{n}+\frac{d\rho}{dt}^{n+1}\times dt,\\
%	\bold{x}_a^{n+1}=\bold{x}_a^{n}+\bold{u}_a^{n+1}\times dt.\\
%	%\end{array}
%\end{eqnarray}
%

Two types of boundary conditions, rigid wall boundary condition (WBC) and periodic boundary condition (PBC), are employed to solve the aforementioned governing equations for fluid flow. In case of WBC, the wavefront hits the side walls, reflects, and generates secondary waves. Since we consider the slip condition, the fluid particles, after interacting with the wall particles, glide parallel to the wall, as shown in the upper panel of Fig. \ref{Fig:PBC}. On the other hand, for PBC, the particle leaves the domain from one side boundary. Another particle comes back into the workspace from the opposite side with the same values of flow parameters (except position), as depicted in the lower panel of Fig. \ref{Fig:PBC}.

\section{Results and discussion}
%Watanable et al., studied dam-break flow with 
%using Lattice Boltzmann Method (LBM) \cite{Watanabe2021}. In article \cite{mu2023}, the authors studied dam-break flow around cylindrical and square pillars using Direct Numerical Simulation (DNS) in OpenFOAM. The impact on the square pillar was studied by varying its angle of the face with the direction of flow and examined the resulting pressure distribution on the pillar. Aureli et al., \cite{ozmen2022exp} studied dambreak flow in a 2D depth-averaged model Finite Volume Method (FVM), a 3D Eulerian two-phase model  Volume of Fluid (VOF). 
%and a 3D Smoothed Particle Hydrodynamics (SPH) model, to estimate the impact load exerted by a dam-break wave on an obstacle. The results were compare with the experiment. 
%Hein et al., studied flow impact on a building array using Large Eddy Simulation (LES) method. Force exerted on the buildings are compared with a 2d FVM and 3d RANs models \cite{hien2021}. There are plenty of numerical results available with the above discussed numerical methods \cite{issakhov2018numerical,ZHANG2018}.
The dam-break is modelled as a column of fluid with dimensions (fluid width $L_y$) $\times$ (fluid length $L_f$) $\times$ (fluid height $H$) initially reserved behind a dam. The dam gate is suddenly removed, ie, water is allowed to move in one direction (say $x$), and the fluid rushes out under gravity, spreading into the downstream region or flowing through a channel with impenetrable walls.

The flow of water after a dam-break is simulated using a rigid wall plus slip boundary condition (WBC) and a periodic boundary condition (PBC) to study their effects on the stopping wall in the downstream, with and without an obstacle in front of the wall. We have considered a channel (workspace) of length $L_x=5.0$m with an initial liquid column of width $L_y=1.0$m. We consider pure water as a fluid with properties, density $(\rho):1000$ kg/$m^3$ and dynamic viscosity $(\mu):0.001$Pa.s. The length of the fluid column is $L_f=0.5$m.
\begin{figure}[H]
	\centering
	\includegraphics[width=0.9\textwidth,height=0.9\textwidth]{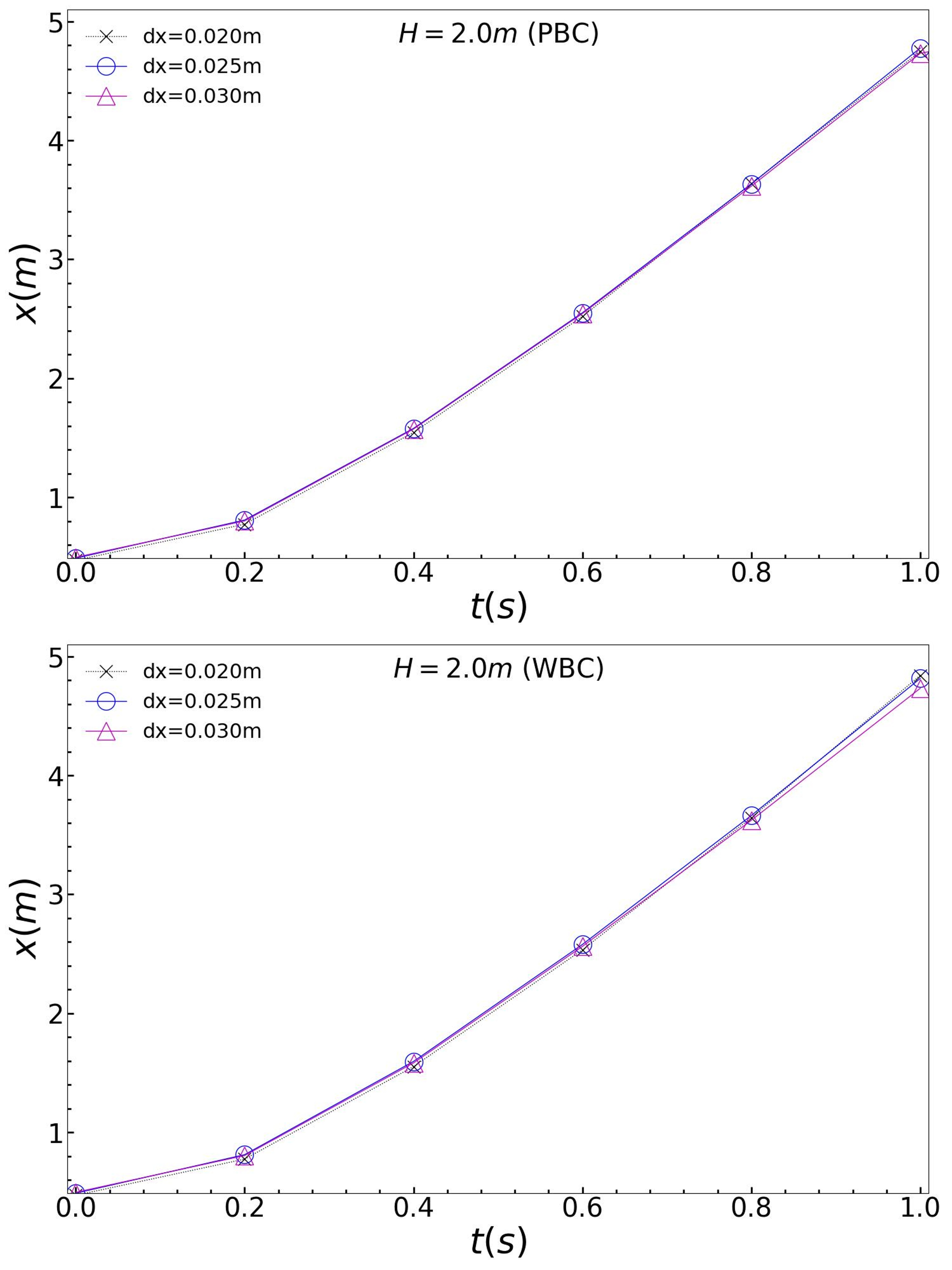}
	\caption{Convergence test with bore front movement for different particle spacing $dx$.}
	\label{Fig:convergence}
\end{figure}
%Fig2
The convergence of our SPH code is tested by tracking the bore-head motion in a plain dam-break confined flow for different $dx(=dy)$ values representing consideration of different numbers of SPH particles ($N$). 
It is observed from Fig. \ref{Fig:convergence} that the curves of the displacements of bore-heads for $dx =0.030m, \ 0.025m,$ and $0.020$m, almost coincide with each other for both the WBC and PBC conditions. We consider $dx = 0.025$m as the initial particle spacing throughout this article. 
\begin{figure}[H]
	\centering
	\includegraphics[width=1\textwidth]{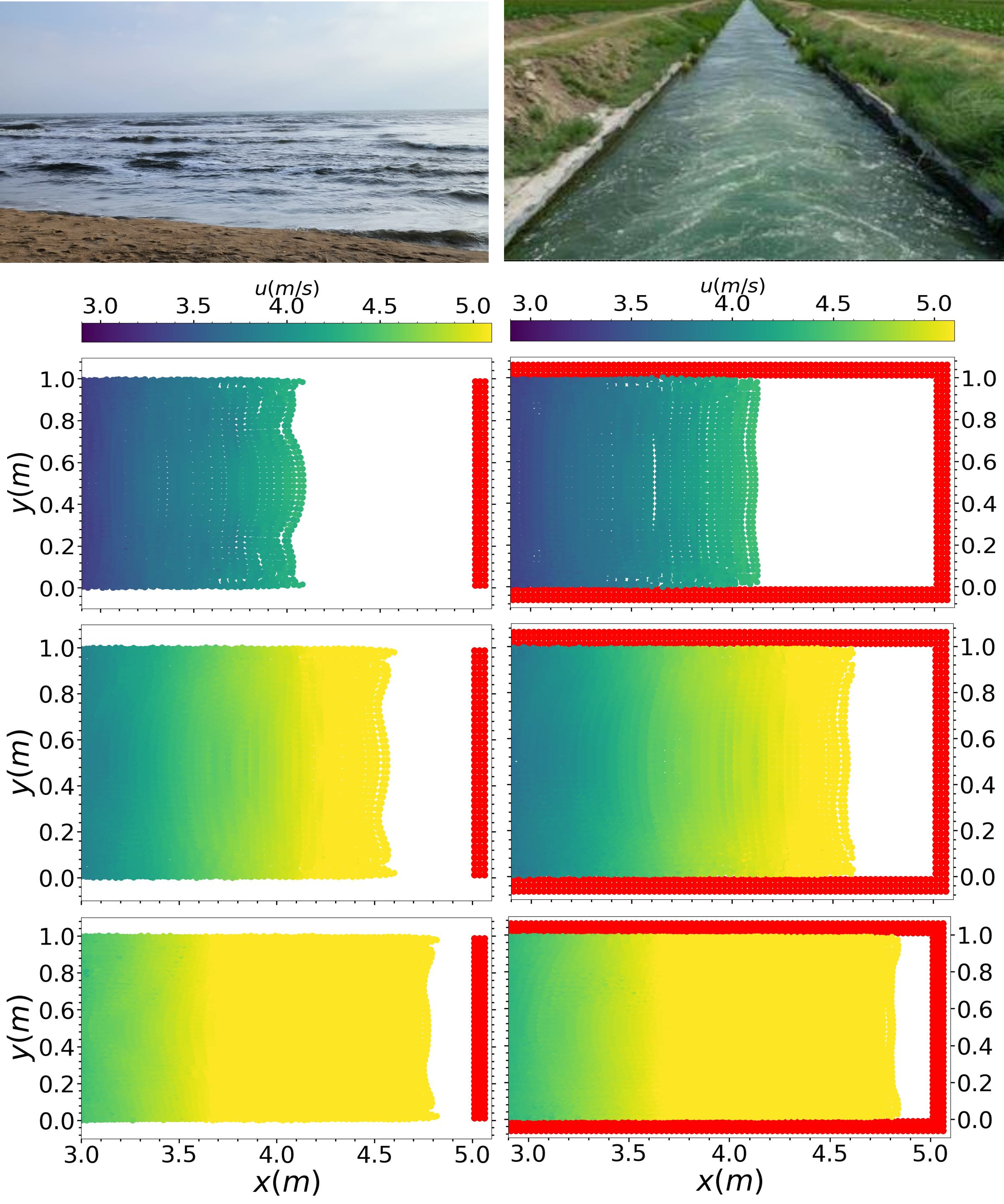}
	\caption{Flow pattern with PBC (first column) and WBC (second column) for dam heights $H=1.0$m,$ 1.5$m, and $2.0$m, respectively, at time $t=1.0$s.}
	\label{Fig:plain}
\end{figure} 
%Fig3
Figure \ref{Fig:plain} presents a top-view of plain dam-break with PBC (left column) and WBC (right column) at an arbitrary time, $t = 1.0$s for dam heights $H = 1.0$m, $1.5$m, and $2.0$m. In the PBC setup, fluid is free to extend beyond the domain edges and is considered to have periodicity along the perpendicular direction to the flow of water. Whereas, in the case of WBC, the flow is confined between the rigid side walls. This confinement has a significant impact on flow dynamics, which yields a difference in the flow pattern of PBC and WBC, and hence the pressure distribution on the wall.  The WBC exhibits almost a plain wavefront when the slip condition is applied. For PBC,  parabolic structures are visible, which are more prominent for a shorter dam height ($H$), i.e., when the flow is slow.
%
% Tab1
Maximum pressure, $P_{max}$ (in Pascal), at a few points on the wall 
($y=0.5$m, $z=0.1$m, $0.2$m, $0.3$m) for dam-break-flows with different $H$ are presented for both  $P_{max}^{PBC}$ and $P_{max}^{WBC}$, in Table \ref{tab:PlainDam}.
The pressure at each point increases with increasing $H$, which is obvious as the taller the dam, the larger the kinetic energy in the dam-break flow. On the other hand, for a specific dam height $H$, $P_{max}$ is large at the bottom of the wall, which decreases with increasing $z$ and becomes zero beyond a certain $z$, where the flow-water can not reach. Moreover, it is observed for a particular $H$, $P_{max}^{WBC}$ are greater than $P_{max}^{PBC}$ as observed   which is very prominent at $z=0.1$m ($P_{max}^{WBC} \approx 2P_{max}^{PBC}$).      
\begin{table}[H]
	\centering
	\scriptsize
		\caption{Maximum pressure $P_{max}$ (in $Pa$) at various $z$ values on the $y=0.5$m line on the wall are presented for PBC, $P_{max}^{PBC}Pa$ and WBC, $(\textcolor{blue}{P_{max}^{WBC}}Pa)$.}
		\label{tab:PlainDam}
	\begin{tabular}{|c|c|c|c|c|c|}
		\hline
		Type&z(m)& H=1.0m &H=1.5m & H=2.0m & H=2.5m \\
		\hline
		\multirow{3}{*}{Np}&0.1&6667.11 (\textcolor{blue}{6977.60})&15822.79 (\textcolor{blue}{17780.52}) & 29517.69 (\textcolor{blue}{37733.68}) & 33724.90 (\textcolor{blue}{61877.82})\\
		&0.2&2108.72(\textcolor{blue}{2716.89})&4642.19(\textcolor{blue}{6299.25})& 19108.75(\textcolor{blue}{22256.39}) & 21802.00 (\textcolor{blue}{27332.63})\\
		&0.3&706.35
		(\textcolor{blue}{840.24})&3191.89(\textcolor{blue}{3327.17})& 6092.61(\textcolor{blue}{6485.56}) & 15072.16(\textcolor{blue}{16473.51}) \\
		\hline
		\end{tabular}
	\end{table}

\begin{figure}[ht]
\centering
\includegraphics[width=1\textwidth]{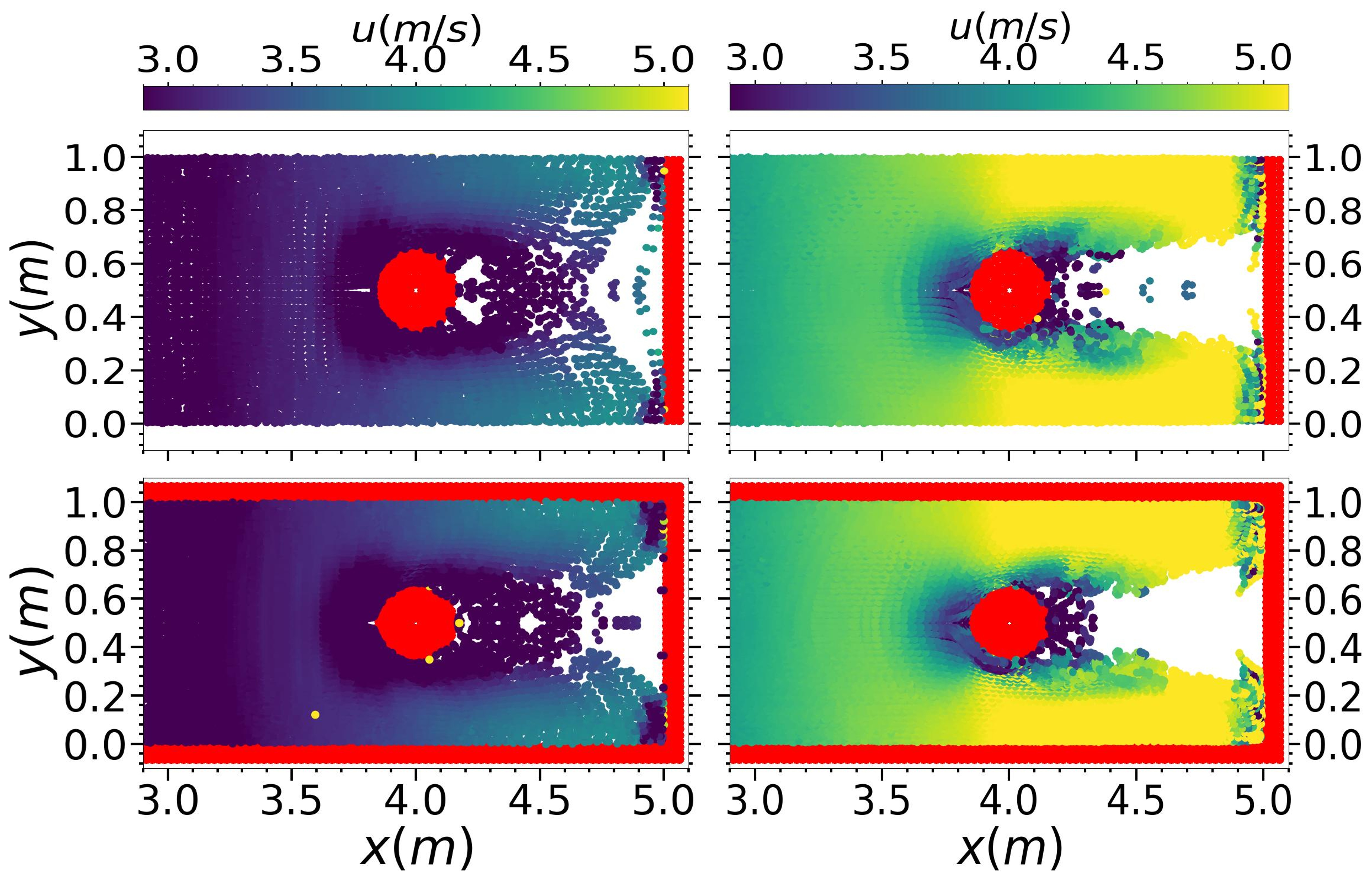}
	\caption{Top ($xy$) view of PBC flow (first row) and WB flow (second row) with pillar $D=0.3$m for $H=1.0$m and $2.0$m.} 
	\label{Fig:dim3height}
\end{figure}

\begin{figure}[H]
	\centering
	\includegraphics[width=1\textwidth]{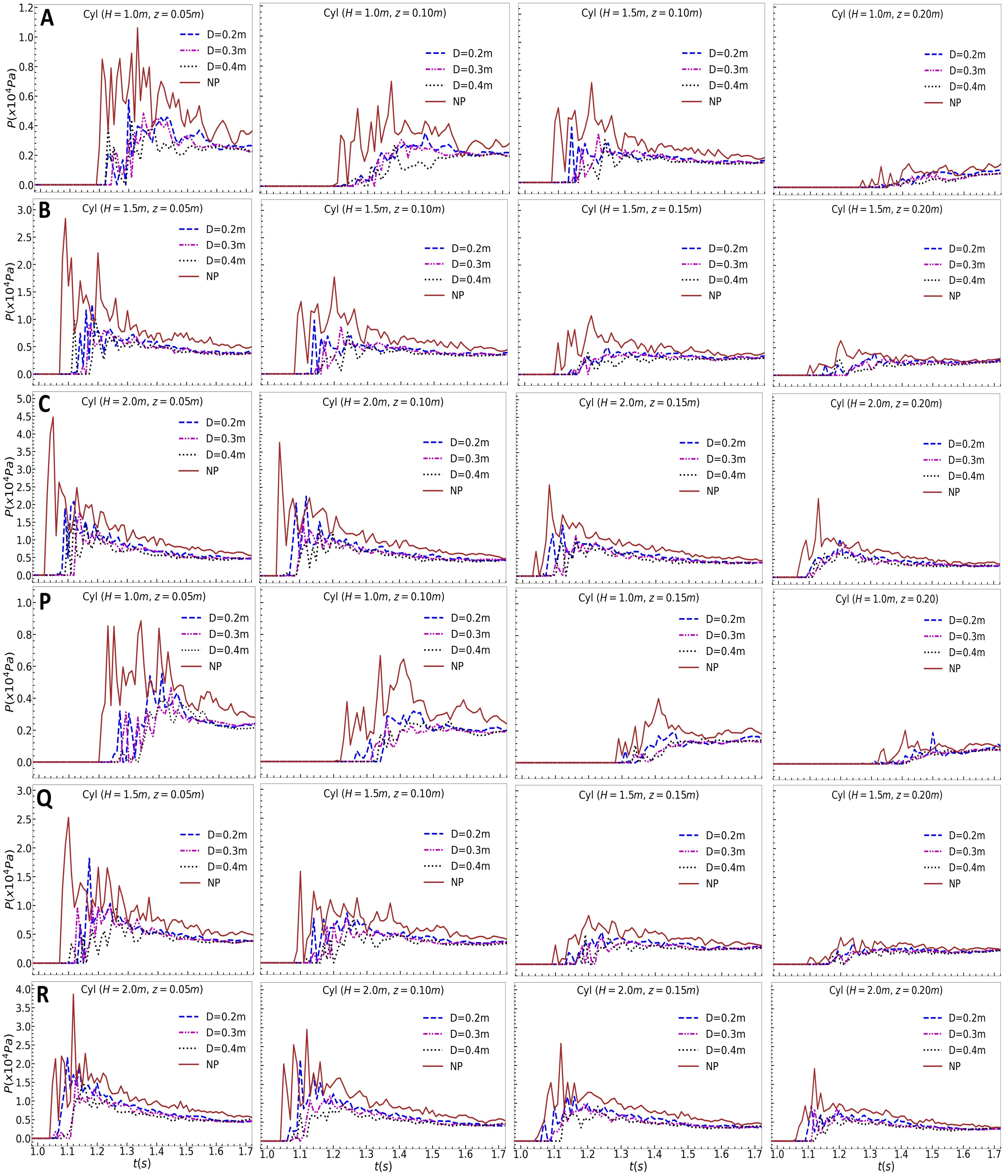}
	\caption{Comparison of pressure on the wall at $y=0.5$m for different cylinder radii for WBC and PBC}
	\label{Fig:PresDiameter}	
\end{figure}

%Fig4
A cylindrical pillar ($D = 0.3$m) placed on the downstream splits the water flow, creating a void portion (shadow) during initial impact on the wall behind it. The no-slip boundary condition is employed for the solid-fluid interaction at the surface of the pillar, a factor that significantly influences the flow dynamics. Particles of the split waves move towards the side boundaries. In the WBC condition at the boundary of the workspace, the particles start moving along the boundary after interacting with the boundary, as we consider the slip condition at the boundary of the workspace. These particles create a shield between the wall and the particles coming from the dam next, and push these newly coming particles towards the shadow. Eventually, the shadow gets filled up. The shadow gets filled up in PBC, too, but the mechanism is different. In this case,  particles of the split flows go out of the working domain after crossing the boundary lines, and particles of the same velocity and acceleration enter the domain from the opposite side boundary, which move towards the shadowed portion as shown in Fig. \ref{Fig:PBC}.
%Fig5
Figure \ref{Fig:dim3height} presents snapshots of the initial impact on the wall for $H = 1.0$m (left panel) and $H = 2.0$m (right panel) for both PBC (top panel) and WBC plus slip (bottom panel). It is evident from the shades (colour) of the contours that the flow velocity is increased for a higher $H$. In comparison to PBC, fluid particles in WBC move more rapidly behind the pillar, leading to a rapid narrowing of the shadow for $H=1.0$m. The velocity distribution and shadow formation are similar in both boundary conditions when the flow speed is large ($H=2.0$m).  
%
%Fig5
An increase in dam height ($H$) results in the storage of higher potential energy, and as the dam-breaks down, water flows with a greater velocity (kinetic energy).
The flowing water hit the pillar ($a=1.0$m, $D=0.3$m),
\begin{figure}[H]
	\centering
%	wall
\includegraphics[width=1\textwidth]{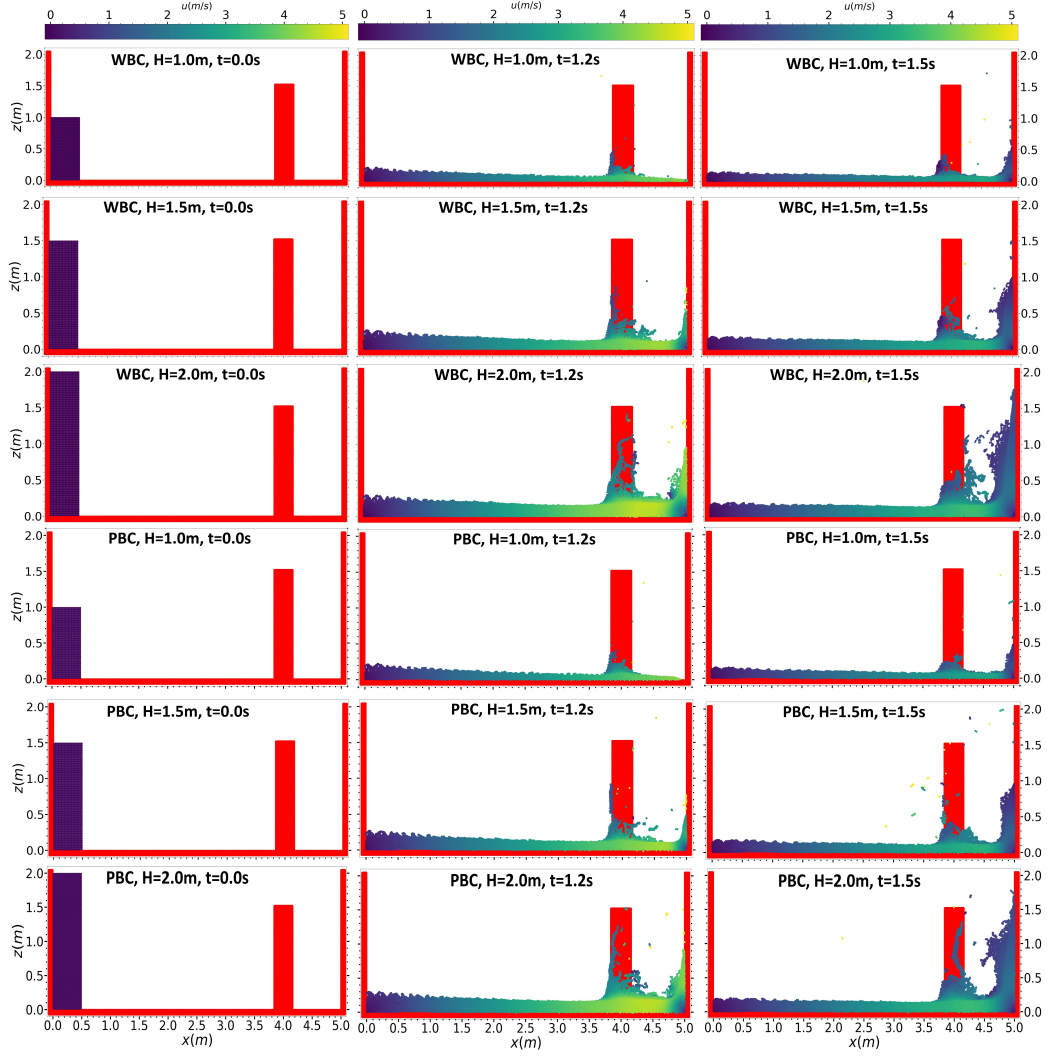}
	\caption{Snapshots of three dimensional dam-break flow. Each row represents fluid column height $H=1.0$m,$1.5$m and $2.0$m.}
	\label{Fig:dynamics}
\end{figure}
\begin{figure}[H]
		\centering
	\includegraphics[width=1\textwidth]{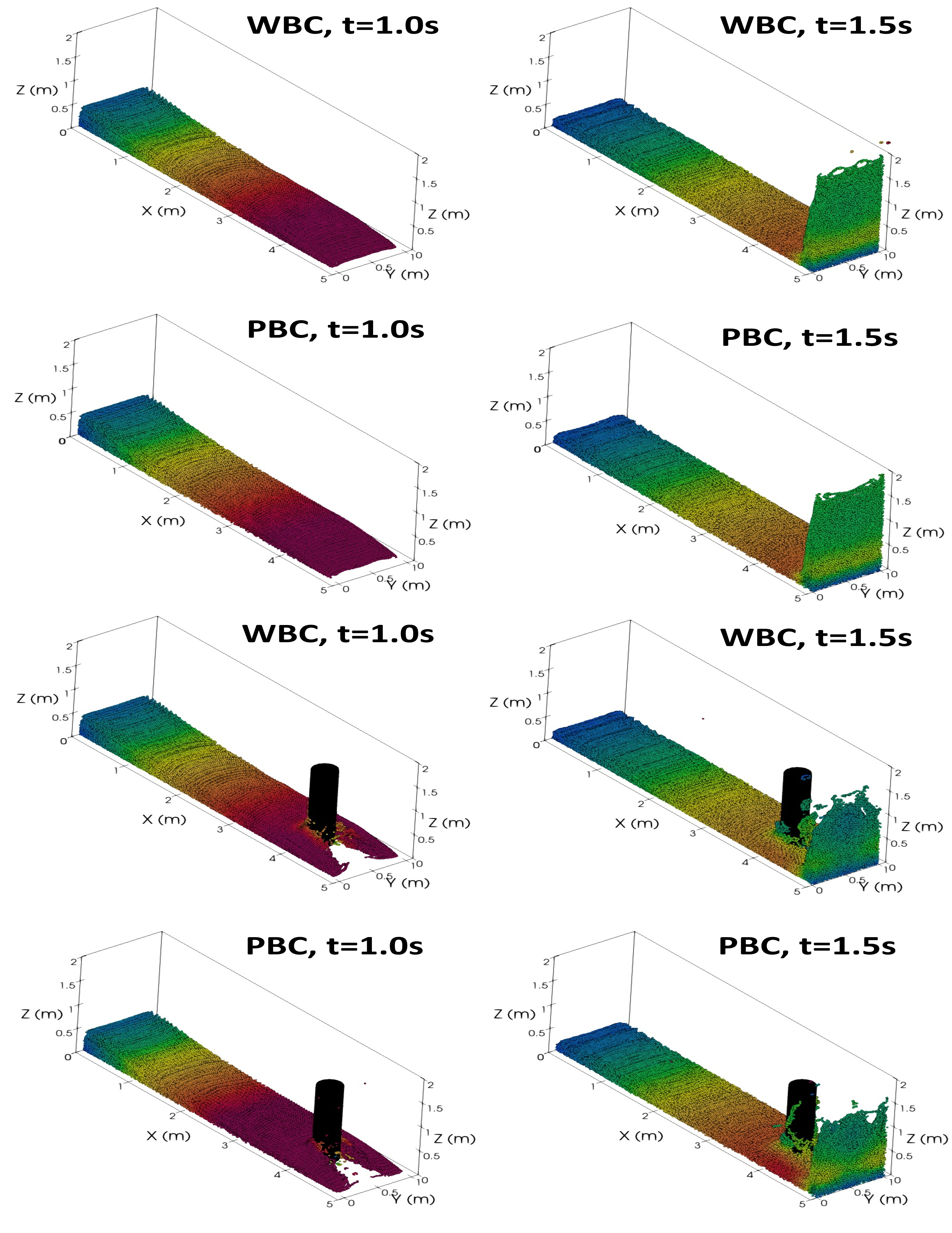}
	\caption{Flow pattern at $t=1.0$s and $1.5$s (1st, and 2nd  column) for dam-height $H=2.0$m. WBC results are shown 1st and 3rd rows whereas PBC cases are in 2nd and 4th rows, respectively.}
	\label{Fig:3d}
\end{figure}
 \noindent bifurcated, and progressed towards the wall behind the pillar, which stopped the fluid motion in the positive $x$-direction. These particles are pushed by the particles coming from the dam next and are compelled to climb up the wall. A few snapshots ($zx$ cross section) at $t = 0.0$s, $1.2$s, and $1.5$s, of the progress of this process are presented in Fig. \ref{Fig:dynamics} for different dam heights. We observe that as the dam height increases,   water climbs higher up along the wall, and more water splashes. The rise in climbing height and splashes is because the taller the dam, the greater is the kinetic energy (velocity) associated with the flow.
 The climbing height and splashes are somewhat higher in WBC in comparison to those obtained from PBC. 
\begin{figure}[H]
	\centering
	\includegraphics[width=1\textwidth,height=0.6\textheight]{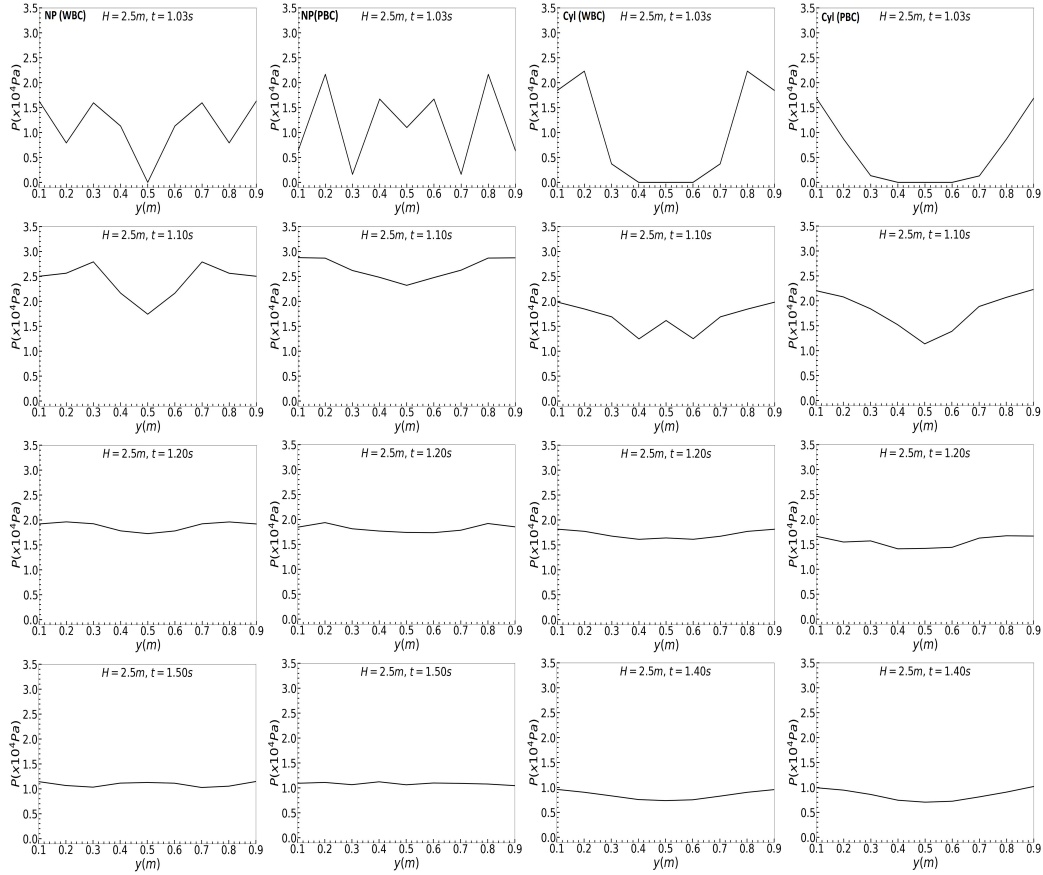}
	\caption{Pressure versus $y$ at $z=0.1$m, $x=5$m on the wall for NP (first two column) and in presence of a cylinder pillar $a=1$m, $D=0.3$m at different time $t$ in WBC and PBC.}
	\label{Fig:surface1}
\end{figure}
\begin{figure}[H]
\includegraphics[width=1\textwidth]{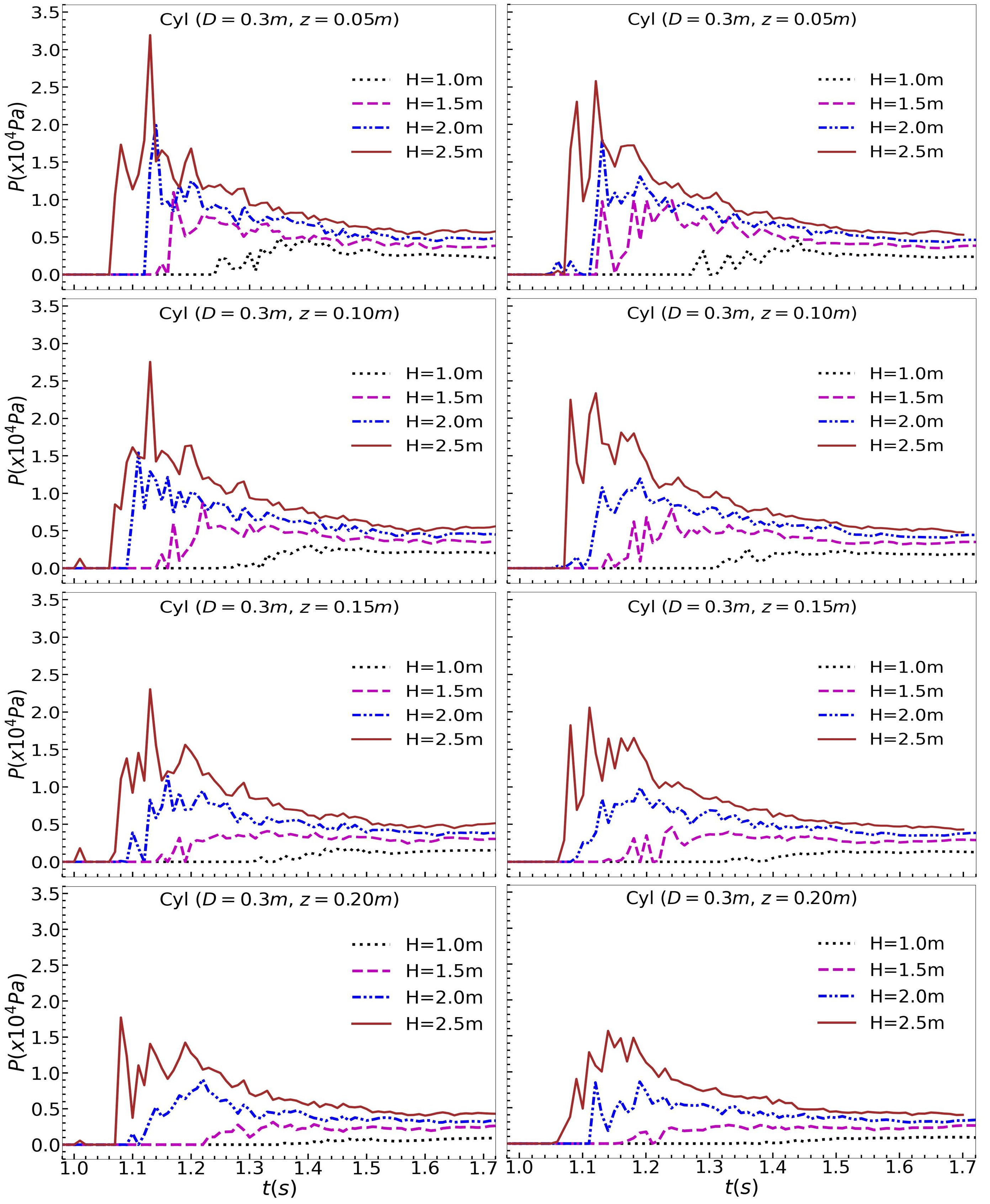}
	\caption{Pressure on the wall with a cylindrical obstacle ( $D=0.3$m) using  WBC and PBC, respectively. The values of pressures presented here are measured on the wall at different $z$ values and $y=0.5$m.}	
	\label{Fig:PresHeightcomp}
\end{figure}
% Fig6
Velocity contour of three-dimensional dam-break flow with and without obstacle (pillar) for $H=2.0$m subjected to WBC and PBC is shown in Fig. \ref{Fig:3d} for $ t=1.0$s and $1.5$s. %{0.0s,0.8s,}  
%plain dam-break flow (first column)  and with cylindrical pillar flow (second row) 
The figure illustrates how the presence of a pillar alters the flow path, causing a noticeable deviation in height when climbing along the wall, more clearly than in the previous figure (Fig. \ref{Fig:dynamics}). The figure demonstrates that the presence of an obstacle reduces flow energy.  As a result, the flow exhibits a reduced ascent along the wall height compared to the dam-break scenario with no pillar. The impact (ascent) of dam-break flows is reduced in cases of PBC compared to WBC.
%
%Fig7
Pressure ($P$) exerted on the down stream wall by the flowing water at $z=0.1$m across the width ($y$ variation) for $H=2.5$m with 
no-obstruction and a cylindrical pillar ($a=1.0$m, $D=0.3$m) as obstacle at different time, $t$ ($1.03$s, $1.10$s and $1.50$s) are displayed in Fig. \ref{Fig:surface1}.
The early-stage impact yields parabolic $P$ versus $y$ curves with $P\rightarrow 0$ in the middle portion. This low $P$ zone is wider in PBC than WBC. 
Structures in the curve of WBC are arising as the water particles are confined within the domain. The curves get flattened with time, and the value of $P$ is also reduced. 
\begin{figure}[H]
	\centering
\centering
\includegraphics[width=1\textwidth]{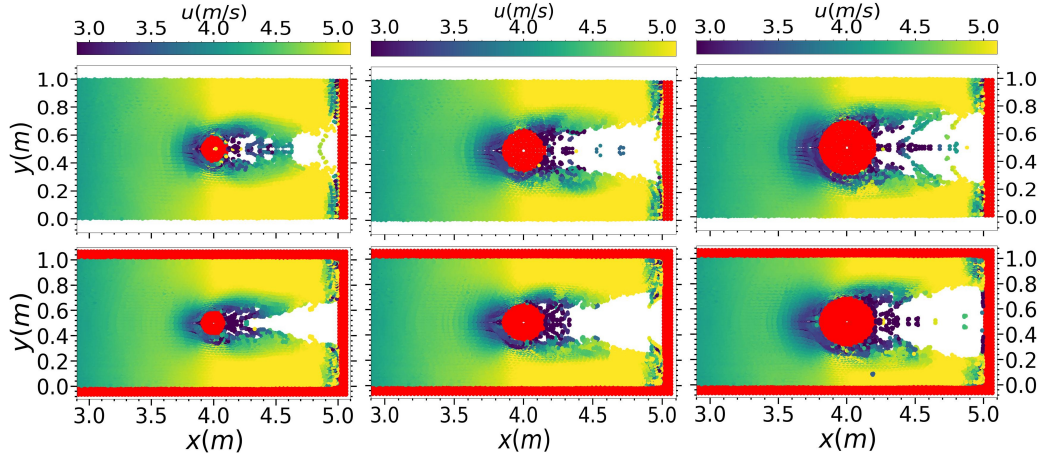}
	\caption{Top view of flow separation and shadow formation for cylindrical obstacles of diameters, $D=0.2$m, $0.3$m and $0.4$m. Both PBC (first row) and WBC (second row) are considered.} 
	\label{Fig:cyl}
\end{figure}
%
% Fig8
Variation of pressures with time,  at the center-line of the wall at $y = 0.5$m for various locations varying $z$ in presence of a cylindrical pillar with diameters $D = 0.3$m  for both WBC (left column) and PBC (right column) flow cases are displayed in Fig. \ref{Fig:PresHeightcomp} for various dam height ($H$). It is noted that the pressure curves are grossly similar for both WBC and PBC, where the pressure in WBC is slightly higher than in PBC. 
It is observed that the highest pressure on the wall occurs for $H = 2.5$m,  while the lowest pressure is recorded for $H = 1.0$m as expected.
%The peak pressure values are similar for PBC and WBC cases.\\
 A close look into  Fig. \ref{Fig:PresHeightcomp} reveals that the $P$ decreases with an increase in $z$. As for the temporal evolution, the maximum pressure occurs a little after (between $t=1.1$s and $1.2$s) when water first makes contact with the wall (at the bottom of the wall). When water starts climbing up, the pressure is reduced. A complex flow is generated by the mixing of the incoming flow from the dam and the water that falls down, which ascends along the wall. This complex flow gives rise to several spikes on the pressure versus time curves. More spikes are observed for a taller dam.
We examine the effects of the obstacle's size on the fluid flow and its impact on the wall behind it. We consider a dam of height $2.0$m, and cylindrical pillars with different diameters ($D = 0.2$m, $0.3$m, and $0.4$m) located at $a = 1.0$m as obstacles. It is observed from Fig. \ref{Fig:cyl} that the flow separation is larger for a larger  $D$, forming a wider shadow, which means a wider zone behind the pillar is protected. The shadow structures in the case of WBC are somewhat larger than those obtained from PBC.
%An increase in $D$ leads to greater flow deflection around the pillar, which affects the pressure distribution on the wall. Comparing the PBC and WBC cases, it is observed that the number of impacted SPH particles is higher in WBC than in PBC, indicating stronger interactions at the boundary.
% Fig10 
In Fig. \ref{Fig:PresDiameter}, the pressure ($P$) on the vertical line at $y=0.5$m on the wall with the change of time is displayed for the cylindrical pillars of hydraulic diameters, $D = 0.2$m, $0.3$m and $0.4$m. The corresponding no-pillar (NP) dam-break data are also presented for comparison. We have considered both WBC and PBC in this calculation, and they are presented in rows A-C and rows P-R, respectively. It is observed that in all cases, $P$ for NP is significantly higher compared to those with obstacles. As the $D$ increases, $P$ decreases. In all cases, the $P$ vs $t$ curve exhibits a peaked structure, where the peaks appear after a certain time from the initial impact of the fluid on the wall. The peaks are occurring earlier for the larger $H$.   Many spikes are visible in all the curves, which arise due to the complex flow dynamics resulting from the interaction of incoming and returning flows. The pressure dissipates for all cases over time and ultimately settles to a very low value. It is evident from the graphs in each row that the pressure ($P$) for NP and all diameter cases decreases and eventually becomes insignificant (flat) as $z$ increases. The trends of the graphs from  WBC are similar to those from PBC but somewhat larger in magnitude. 
% Tab2
\begin{table}[ht]
	\centering
	%	\small
	\scriptsize
	\caption{$P_{max}^{PBC}$ ($\textcolor{blue}{P_{max}^{WBC}}$) in Pascal on the stopping wall at ($y=0.5$m,$z$) for cylindrical pillars of  different $D$ positioned at $a=1.0$m. Data in the second rows (in italics) for each $z$ value represent the percentage of reduction of $P_{max}$ with respect to the corresponding NP cases illustrated in Table \ref{tab:PlainDam}. 
	}
	\label{Tab:PeakDiam1}
		\begin{tabular}{|c|c|c|c|c|c|}
		\hline
		Type&z(m)& H=1.0m &H=1.5m & H=2.0m & H=2.5m \\
		\hline			
		\multirow{3}{*}{D=0.2m }&\multirow{2}{*}{0.1}&3166.78(\textcolor{blue}{3544.23})&8595.77(\textcolor{blue}{9823.34})& 21520.46(\textcolor{blue}{22466.94})& 28888.87(\textcolor{blue}{30105.03}) \\
		&&\it 52.50 (\textcolor{blue}{49.21}) & \it 45.67 (\textcolor{blue}{44.75}) &\it 27.09 (\textcolor{blue}{40.46}) &\it 14.34 (\textcolor{blue}{51.35})\\
		\cline{2-6}			 			
		&\multirow{2}{*}{0.2}&1962.64(\textcolor{blue}{2269.79})&3611.41(\textcolor{blue}{4231.01})& 9307.03(\textcolor{blue}{10869.78}) &18380.83(\textcolor{blue}{23561.64}) \\
		&&\it 6.93  (\textcolor{blue}{16.46}) &\it 22.20 (\textcolor{blue}{32.83}) &\it 51.29 (\textcolor{blue}{51.16}) &\it 15.69  (\textcolor{blue}{13.80}) \\
		\cline{2-6}
		&\multirow{2}{*}{0.3}&510.10(\textcolor{blue}{567.01})& 2521.60(\textcolor{blue}{3236.47})&4988.65(\textcolor{blue}{5563.64}) &9913.67(\textcolor{blue}{10594.30}) \\
		&&\it27.78 (\textcolor{blue}{32.52}) &\it 21.00 (\textcolor{blue}{2.73})  &\it 18.12 (\textcolor{blue}{14.21}) &\it 34.23 (\textcolor{blue}{35.69}) \\
		\hline
		\multirow{3}{*}{D=0.3m }&\multirow{2}{*}{0.1}&2629.68(\textcolor{blue}{3092.97})&7900.00(\textcolor{blue}{8728.11})& 11955.74(\textcolor{blue}{15425.40})& 23352.85(\textcolor{blue}{27552.42}) \\
		&&\it 60.56 (\textcolor{blue}{55.67}) &\it 50.07 (\textcolor{blue}{50.91}) &\it 59.50 (\textcolor{blue}{59.12}) &\it 30.75 (\textcolor{blue}{55.47}) \\
		\cline{2-6}
		&\multirow{2}{*}{0.2}&1125.38(\textcolor{blue}{1145.18})&2581.18(\textcolor{blue}{3143.91})& 8609.21(\textcolor{blue}{9091.38}) &15689.30(\textcolor{blue}{17712.95}) \\
		&&\it46.63 (\textcolor{blue}{57.85}) &\it 44.40 (\textcolor{blue}{50.09}) &\it 54.95 (\textcolor{blue}{59.15}) &\it 28.00 (\textcolor{blue}{35.19}) \\
		\cline{2-6}
		&\multirow{2}{*}{0.3}&417.59(\textcolor{blue}{477.60})& 1961.81(\textcolor{blue}{2057.43})&4758.32(\textcolor{blue}{5010.19}) &13064.88(\textcolor{blue}{14541.99}) \\
		&&\it 40.88 (\textcolor{blue}{43.16}) &\it 38.54 (\textcolor{blue}{38.16}) &\it 21.90 (\textcolor{blue}{22.75}) &\it 13.32 (\textcolor{blue}{11.73}) \\
		\hline
		\multirow{3}{*}{D=0.4m }&\multirow{2}{*}{0.1}&
		2505.44(\textcolor{blue}{2755.83})&7581.06(\textcolor{blue}{7621.35})&9639.39 (\textcolor{blue}{13294.15})& 18050.99(\textcolor{blue}{21272.95}) \\
		& &\it 62.42 (\textcolor{blue}{60.50}) &\it 52.09 (\textcolor{blue}{57.14}) &\it 67.34 (\textcolor{blue}{64.77}) &\it 46.48 (\textcolor{blue}{65.62}) \\
		\cline{2-6}
		&\multirow{2}{*}{0.2}&885.77(\textcolor{blue}{1105.51})&2604.74(\textcolor{blue}{2874.55})&6621.45 (\textcolor{blue}{6727.02}) &14479.91(\textcolor{blue}{15847.45}) \\
		&&\it 57.99 (\textcolor{blue}{67.00}) &\it 43.89 (\textcolor{blue}{54.37}) &\it 65.35 (\textcolor{blue}{69.77}) &\it 33.58 (\textcolor{blue}{42.02}) \\
		\cline{2-6}
		&\multirow{2}{*}{0.3}& 210.05(\textcolor{blue}{370.82})& 1842.80 (\textcolor{blue}{1973.54})& 3959.06(\textcolor{blue}{3966.99}) & 10819.07(\textcolor{blue}{11247.58}) \\
		&&\it 70.26 (\textcolor{blue}{55.87}) &\it 42.27 (\textcolor{blue}{40.68}) &\it 35.02 (\textcolor{blue}{38.83}) &\it 28.22 (\textcolor{blue}{31.72}) \\
		\hline
	\end{tabular}
\end{table}

% Tab2
The maximum values of pressure, $P_{max}^{PBC}$  and, $\textcolor{blue}{P_{max}^{WBC}}$ (in Pascal) on the stopping wall at various $z$ along $y=0.5$m for cylindrical pillars of  different hydraulic diameter ($D=0.2$m,$0.3$m, and $0.4$m) placed  at $a=1.0$m are presented in Table \ref{Tab:PeakDiam1}. Data in the second row (italics) for each $z$ value represents the percentage of reduction of $P_{max}$ with respect to the corresponding pressure of NP cases shown in Table \ref{tab:PlainDam}. It is observed that $P_{max}$ at a particular $z$ increases with the dam height for both PBC and WBC, but the increase is not as robust as seen for NP. For a particular $H$ and $D$, $P_{max}$ is decreases with increasing $z$.  At a particular $H$ and $z$, the pressure decreases with the increase in $D$. The second row for each $D$ and $z$ (in italics) represents the percentage of the reduction of the pressure for the presence of the pillar obstacle, compared to the corresponding value for the no pillar case (NP). The percentage of reduction in $P_{max}$ in the presence of a pillar compared to $NP$ ($P_{max}^{NP}$) is mostly by $40\%$ to $60\%$. We do not find any regular pattern in percentage of pressure reduction, because, the spikes on both the curves (NP and PBC/WBC, the difference of which are considered to calculate the percentage of pressure reduction, $\epsilon=\{(P_{max}^{NP}-P_{max})/P_{max}^{NP}\}\times 100$ occur randomly. This mismatch in spikes gives rise to such a random pressure reduction locally.

% Fig11
\begin{figure}[H]
	\centering
	\includegraphics[width=1\textwidth]{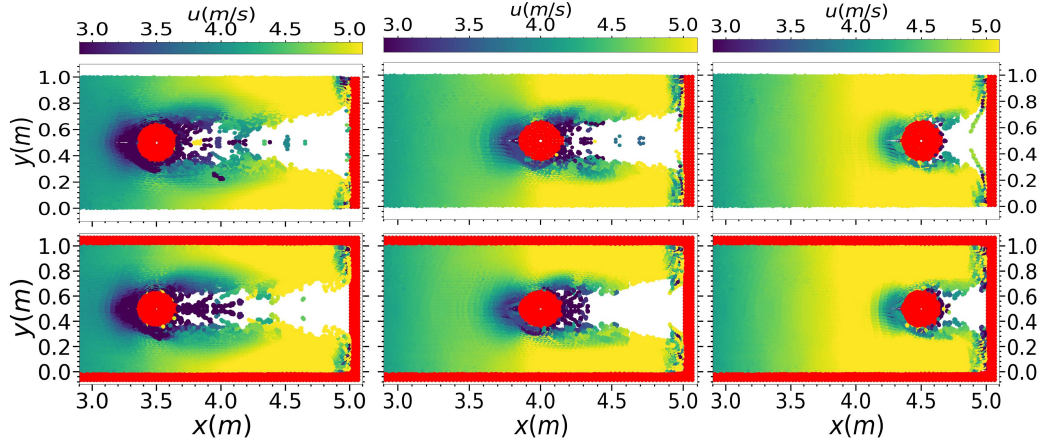}
	\caption{Top view of three dimensional dam-break flow with $D=0.3$m and $H=2.0$m. First row shows various pillar position ($a=1.5$m, $1.0$m, and $0.5$m) for PBC. Second row displays results for WBC.} 
	\label{Fig:cylpos}
\end{figure}
\begin{figure}[H]
\centering
\includegraphics[width=1\textwidth]{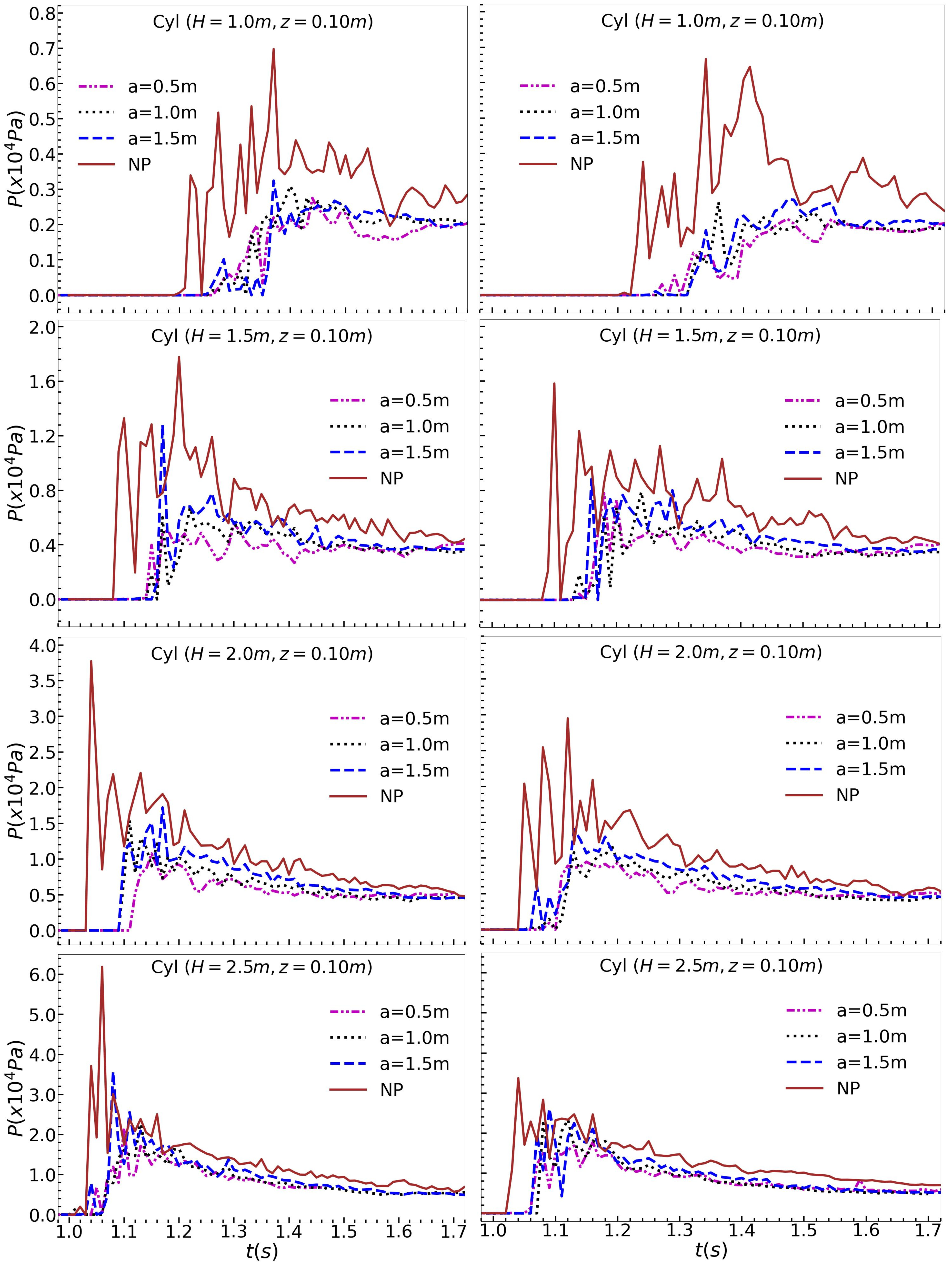}
	\caption{Pressure on the dam wall on pillar positions $a=0.5$m, $1.0$m and $1.5$m for cylindrical pillar of $d=0.3$m. The left figure represents WBC, and the right is PBC.}
	\label{PresPillarPosComp}
\end{figure}

% Fig11
The effects of the location ($a = 0.5$m, $1.0$m, and $1.5$m) of the obstructing pillar ($D=0.3$m) on the dam-break ($H=2.0$m) flow pattern and its impact on the wall behind it, as displayed by snap shots at time of initial impact of water with the wall in Fig. \ref{Fig:cylpos}, have practical implications for  safety of structures in dam-break flow. For both PBC and WBC, the shadowed portions behind the pillar get changed with $a$. For a larger $a$, water particles from the split flows return to the central zone of the channel from the boundary lines, reducing the shadow and hence increasing the impact on the wall.
A pillar, close to the wall (say, $a = 0.5$m), reduces the chance of a merger of the split flow. The shadowed portion in WBC shows a periodic structure, whereas the shadows from PBC are triangles with straight sides. 
%Fig12
Variations of $P_{max}$ on the wall at ($y=0.5$m, $z=0.1$m) with time when a cylindrical pillar ($D=0.3$m) placed at $a$ ($= 0.5$m, $1.0$m, and $1.5$m) is presented in Fig. \ref{PresPillarPosComp}, along with the same from NP condition.   Each row in the figure represents results with different dam heights $H$. WBC  and PBC results are displayed in the left column and right column, respectively. The overall results indicate similar flow behaviour for both boundary conditions, with the key distinction being that, in the case of $P_{max}$, the peak pressure is higher for WBC than PBC. The maximum pressure for the flow with NP  consistently dominates over all cases with pillar for all dam heights $H$.
\noindent Flow with the pillar far from the wall (say,$a = 1.5$m) exhibits an increased pressure value compared to those positioned closer to the stopping wall. This may be attributed to the early merger of the split flows, resulting from the pillar in the shadowed portion, giving rise to localised pressure amplification.
The least pressure is experienced for $a = 0.5$m among the cases considered here. 
% Tab3
The values of $P_{max}^{PBC}$ ($\textcolor{blue}{P_{max}^{WBC}}$) on the wall at various $z$ along $y=0.5$m for cylindrical pillars of $D=0.3$m placed at different locations $a=0.5$m, $1.0$m  and $1.5$m are displayed in Table \ref{Tab:PeakPos}.  
We observe that $P_{max}^{WBC}$ is greater than the corresponding $P_{max}^{PBC}$ for all cases shown here.
We also notice that both $P_{max}^{PBC}$ and $P_{max}^{WBC}$  at a particular $z$ increase with the increase in $H$ for all $a$. For a particular $H$ and $a$, $P_{max}$ decreases with increasing $z$.  At a particular $H$ and $z$, the pressure increases with the increase in $a$.
The maximum percentage of reduction in $P_{max}$ in the presence of a pillar compared to the corresponding  $NP$ values ($P_{max}^{NP}$) is about $70\%$ for the cases considered here.

\begin{figure}[H]
	\centering
	\includegraphics[width=1\textwidth]{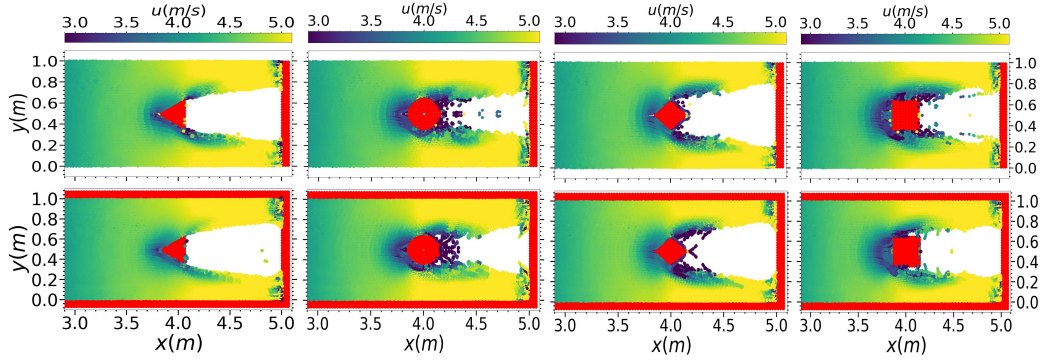}
	\caption{The shape variation of pillars for PBC flow (first row) and WB flow (second row) for $H=2.0$m.} 
	\label{Fig:Shape}
\end{figure}

\begin{table}[H]
	\centering
	%	\small
	\scriptsize
	\caption{Maximum pressure value at $y=0.5$m for a cylindrical pillar ($D=0.3$m) are displayed. The data is given in the format $P_{PBC} (\textcolor{blue}{P_{WBC}})$, where $P_{PBC}$ represents the peak pressure under periodic boundary conditions, and $\textcolor{blue}{P_{WBC}}$ denotes the peak pressure in WBC.}
	%\label{Tab:PeakPres}
	\label{Tab:PeakPos}
	\begin{tabular}{|c|c|c|c|c|c|}
		\hline
		Type&z(m)& H=1.0m &H=1.5m & H=2.0m & H=2.5m \\
		\hline			
		%	\multirow{3}{*}{Np}&0.1&6667.11 (\textcolor{blue}{6977.60})&15822.79 (\textcolor{blue}{17780.52}) & 29517.69 (\textcolor{blue}{37733.68}) & 33724.90 (\textcolor{blue}{61877.82})\\
		%	&0.2&2108.72(\textcolor{blue}{2716.89})&4642.19(\textcolor{blue}{6299.25})& 19108.75(\textcolor{blue}{22256.39}) & 19802.0(\textcolor{blue}{27332.63})\\
		%	&0.3&706.35
		%	(\textcolor{blue}{840.24})&3191.89(\textcolor{blue}{3327.17})& 6092.61(\textcolor{blue}{6485.56}) & 15072.16(\textcolor{blue}{16473.51}) \\
		%	\hline
		\multirow{3}{*}{a=1.5m }&0.1&2706.88(\textcolor{blue}{3241.62})&8843.89(\textcolor{blue}{12801.85})& 14021.47(\textcolor{blue}{17212.85})& 26114.66(\textcolor{blue}{35662.59}) \\
%		&  &\it 59.40 (\textcolor{blue}{53.54}) &\it 44.11 (\textcolor{blue}{28.00}) &\it 52.50 (\textcolor{blue}{54.38}) &\it 22.57 (\textcolor{blue}{42.37}) \\
\cline{2-6}
		&0.2&1559.10(\textcolor{blue}{1689.97})&3181.23(\textcolor{blue}{3444.78})& 10484.07(\textcolor{blue}{11482.98})&18656.29 (\textcolor{blue}{24092.45}) \\
%		&  &\it 26.06 (\textcolor{blue}{37.80}) &\it 31.47 (\textcolor{blue}{45.31}) &\it 45.13 (\textcolor{blue}{48.41}) &\it 5.79  (\textcolor{blue}{11.85}) \\
\cline{2-6}
		&0.3&673.94(\textcolor{blue}{711.01})&2181.79(\textcolor{blue}{2236.19})& 4695.03(\textcolor{blue}{5688.99})& 14026.52(\textcolor{blue}{15977.19}) \\
%		&  &\it 4.59  (\textcolor{blue}{15.38}) &\it 31.65 (\textcolor{blue}{32.79}) &\it 22.94 (\textcolor{blue}{12.28}) &\it 6.94  (\textcolor{blue}{3.01}) \\
\cline{2-6}
		\hline
		\multirow{3}{*}{a=1.0m }&0.1&2629.68(\textcolor{blue}{3092.97})&7900.00(\textcolor{blue}{8728.11})& 11955.74(\textcolor{blue}{15425.40})& 23352.85(\textcolor{blue}{27552.42}) \\
%		&  &\it 60.56 (\textcolor{blue}{55.67}) &\it 50.07 (\textcolor{blue}{50.91}) &\it 59.50 (\textcolor{blue}{59.12}) &\it 30.75 (\textcolor{blue}{61.94}) \\
\cline{2-6}
		&0.2&1125.38(\textcolor{blue}{1145.18})&2581.18(\textcolor{blue}{3143.91})& 8609.21(\textcolor{blue}{9091.38}) &15689.30(\textcolor{blue}{17712.95}) \\
%		& &\it 46.63 (\textcolor{blue}{57.85}) &\it 44.40 (\textcolor{blue}{50.09}) &\it 54.95 (\textcolor{blue}{59.15}) &\it 20.77 (\textcolor{blue}{35.19}) \\
\cline{2-6}
		&0.3&417.59(\textcolor{blue}{477.60})& 1961.81(\textcolor{blue}{2057.43})&4758.32(\textcolor{blue}{5010.19}) &13064.88(\textcolor{blue}{14541.99}) \\
%		&  &\it 40.88 (\textcolor{blue}{43.16}) &\it 63.62 (\textcolor{blue}{38.16}) &\it 21.90 (\textcolor{blue}{22.75}) &\it 70.90 (\textcolor{blue}{11.73}) \\
\cline{2-6}
		\hline
		\multirow{3}{*}{a=0.5m }&0.1&2256.88(\textcolor{blue}{2750.78})&7800.00(\textcolor{blue}{5534.24})&9467.02 (\textcolor{blue}{10769.77})&17691.15(\textcolor{blue}{21537.46}) \\
%		&  &\it 66.15 (\textcolor{blue}{60.58}) &\it 50.70 (\textcolor{blue}{68.87}) &\it 67.93 (\textcolor{blue}{71.46}) &\it 47.54 (\textcolor{blue}{65.19}) \\
\cline{2-6}
		&0.2&1075.32
		(\textcolor{blue}{1120.36})&1718.85(\textcolor{blue}{2798.82})& 7842.71(\textcolor{blue}{8070.67})&13889.70 (\textcolor{blue}{17215.26}) \\
%		&  &\it 49.01 (\textcolor{blue}{62.44		}) &\it 62.97 (\textcolor{blue}{55.57}) &\it 58.96 (\textcolor{blue}{63.74}) &\it 29.86 (\textcolor{blue}{37.02}) \\
\cline{2-6}
		&0.3&229.28(\textcolor{blue}{337.87})&1666.61(\textcolor{blue}{1861.27})& 4386.32(\textcolor{blue}{4523.23})& 12623.10(\textcolor{blue}{13768.19}) \\
%		& &\it 67.54 (\textcolor{blue}{59.79}) &\it 47.79 (\textcolor{blue}{44.06}) &\it 28.01 (\textcolor{blue}{30.26}) &\it 16.25 (\textcolor{blue}{16.42}) \\
		\hline
	\end{tabular}
\end{table}
We study the influences of the shape of the obstacle on flow dynamics. We consider pillars of four distinct geometries: cylinder, triangle, rhombus, and square, all maintaining a hydraulic diameter of $D = 0.3$m. The results presented in  Fig. \ref{Fig:Shape} are computed considering $H = 2.0$m and $a = 1.0$m with WBC and PBC for all shapes.
\begin{table}[H]
	\centering
	%	\small
	\scriptsize
	\caption{Maximum pressure value (in $Pa$) at $y=0.5$m at pillar position $a=1.0$m for $D=0.3$m for different pillar shapes. The data is given in the format $P_{PBC} (\textcolor{blue}{P_{WBC}})$, where $P_{PBC}$ is the peak pressure under PBC, and $\textcolor{blue}{P_{WBC}}$ denotes the same in WBC.}
	\label{Tab:PeakPresShape}
	\begin{tabular}{|c|c|c|c|c|c|}
		\hline
		Type&z(m)& H=1.0m &H=1.5m & H=2.0m & H=2.5m \\
		\hline
		\multirow{3}{*}{Cyl}&0.1&2629.68(\textcolor{blue}{3092.97})&7900.00(\textcolor{blue}{8728.11})& 11955.74(\textcolor{blue}{15425.40})& 23352.85(\textcolor{blue}{27552.42}) \\
		%		&&\it 60.56 (\textcolor{blue}{55.67}) &\it 50.07 (\textcolor{blue}{50.91}) &\it 59.50 (\textcolor{blue}{59.12}) &\it 30.75 (\textcolor{blue}{61.94}) \\
		\cline{2-6}
		&0.2&1125.38(\textcolor{blue}{1145.18})&2581.18(\textcolor{blue}{3143.91})& 8609.21(\textcolor{blue}{9091.38}) &15689.30(\textcolor{blue}{17712.95}) \\
		%		&&\it46.63 (\textcolor{blue}{57.85}) &\it 44.40 (\textcolor{blue}{50.09}) &\it 54.95 (\textcolor{blue}{59.15}) &\it 20.77 (\textcolor{blue}{35.19}) \\
		\cline{2-6}
		&0.3&417.59(\textcolor{blue}{477.60})& 1961.81(\textcolor{blue}{2057.43})&4758.32(\textcolor{blue}{5010.19}) &13064.88(\textcolor{blue}{14541.99}) \\
		%		&&\it 40.88 (\textcolor{blue}{43.16}) &\it 38.54 (\textcolor{blue}{38.16}) &\it 21.90 (\textcolor{blue}{22.75}) &\it 13.32 (\textcolor{blue}{11.73}) \\
		\hline
		\multirow{3}{*}{Tr}&0.1&1945.58(\textcolor{blue}{2095.58})&4879.66(\textcolor{blue}{5598.10})&10576.75(\textcolor{blue}{13520.41})& 19819.208(\textcolor{blue}{21676.73})\\
		%		&  &\it 70.82 (\textcolor{blue}{69.97}) &\it 69.16 (\textcolor{blue}{68.52}) &\it 64.17 (\textcolor{blue}{64.17}) &\it 41.23 (\textcolor{blue}{64.97}) \\	
		\cline{2-6}
		&0.2&1090.44(\textcolor{blue}{1109.81})&2346.20(\textcolor{blue}{3056.21})&7718.16(\textcolor{blue}{8514.91})& 14581.978(\textcolor{blue}{15772.18})
		\\
		%		&  &\it 48.29 (\textcolor{blue}{59.15}) &\it 49.46 (\textcolor{blue}{51.48}) &\it 59.61 (\textcolor{blue}{61.74}) & \it 26.36 (\textcolor{blue}{42.30}) \\
		\cline{2-6}
		&0.3&205.54 (\textcolor{blue}{339.32})&1844.49(\textcolor{blue}{1961.00}) & 4419.033(\textcolor{blue}{5523.64})	 &10366.178(\textcolor{blue}{11576.85})
		\\
		%		&  &\it 70.90 (\textcolor{blue}{59.62}) &\it 42.21 (\textcolor{blue}{41.06}) &\it 27.47 (\textcolor{blue}{14.83}) &\it 31.22 (\textcolor{blue}{29.72}) \\
		\hline
		\multirow{3}{*}{Rh}&0.1&1881.93(\textcolor{blue}{1896.07})&4569.27(\textcolor{blue}{5245.10
		})&8951.72(\textcolor{blue}{12209.51})&18500.00(\textcolor{blue}{19022.65})
		\\
		%		& &\it 71.77 (\textcolor{blue}{72.83}) &\it 71.12 (\textcolor{blue}{70.50}) &\it 69.67 (\textcolor{blue}{67.64}) &\it 45.14 (\textcolor{blue}{69.26}) \\
		\cline{2-6}
		&0.2&915.28(\textcolor{blue}{1066.97})&2970.45(\textcolor{blue}{2990.80})&7108.99(\textcolor{blue}{7241.79})&13429.28(\textcolor{blue}{14096.97}) \\
		%		&  &\it 56.60 (\textcolor{blue}{60.73}) &\it 36.01 (\textcolor{blue}{52.52}) &\it 62.80 (\textcolor{blue}{67.46}) &\it 32.18 (\textcolor{blue}{48.42}) \\
		\cline{2-6}
		&0.3&188.98(\textcolor{blue}{506.92}) & 1794.14(\textcolor{blue}{1821.84})&3409.17(\textcolor{blue}{5216.00})&8775.92(\textcolor{blue}{9824.29}) \\
		%		& &\it 73.25 (\textcolor{blue}{39.67}) &\it 43.79 (\textcolor{blue}{45.24}) &\it 44.04 (\textcolor{blue}{19.58}) &\it 41.77 (\textcolor{blue}{40.36}) \\	\cline{2-6}
		\hline			
		\multirow{3}{*}{Sq}&0.1&1688.10(\textcolor{blue}{1722.72})&3892.03(\textcolor{blue}{4526.28})&8611.20(\textcolor{blue}{10977.81})&16738.78(\textcolor{blue}{17986.33}) \\
		%		&  &\it 73.22 (\textcolor{blue}{73.90}) &\it 75.40 (\textcolor{blue}{74.54}) &\it 70.83 (\textcolor{blue}{70.91}) &\it 50.37 (\textcolor{blue}{70.93}) \\	
		\cline{2-6}
		&0.2&692.09(\textcolor{blue}{975.99})&1850.63(\textcolor{blue}{2487.62})& 6590.02(\textcolor{blue}{6847.84})& 12023.37(\textcolor{blue}{13510.26
		}) \\
		%		&  &\it 67.18 (\textcolor{blue}{64.08}) &\it 60.13 (\textcolor{blue}{60.51}) &\it 65.51 (\textcolor{blue}{69.23}) &\it 39.28 (\textcolor{blue}{50.57}) \\	
		\cline{2-6}
		&0.3&102.17(\textcolor{blue}{302.90})&1527.27(\textcolor{blue}{1603.10})&2996.62(\textcolor{blue}{4196.51}) &7002.06(\textcolor{blue}{11914.09
		})\\
		%		& &\it 85.54 (\textcolor{blue}{63.95}) &\it 52.15 (\textcolor{blue}{51.82}) &\it 50.82 (\textcolor{blue}{35.29}) &\it 53.54 (\textcolor{blue}{27.68}) \\
		\hline
	\end{tabular}		
\end{table}
It is noted from the velocity profiles (at the instant of first impact of the flow of water on the wall) displayed in Fig. \ref{Fig:Shape} that the shape and size of the shadows are significantly changed with the change of the shape of the obstacle. It is observed from the velocity contours that velocity reduction is maximum in front of the square and cylindrical shapes, respectively. 
\begin{figure}[H]
	\centering
\includegraphics[width=1\textwidth]{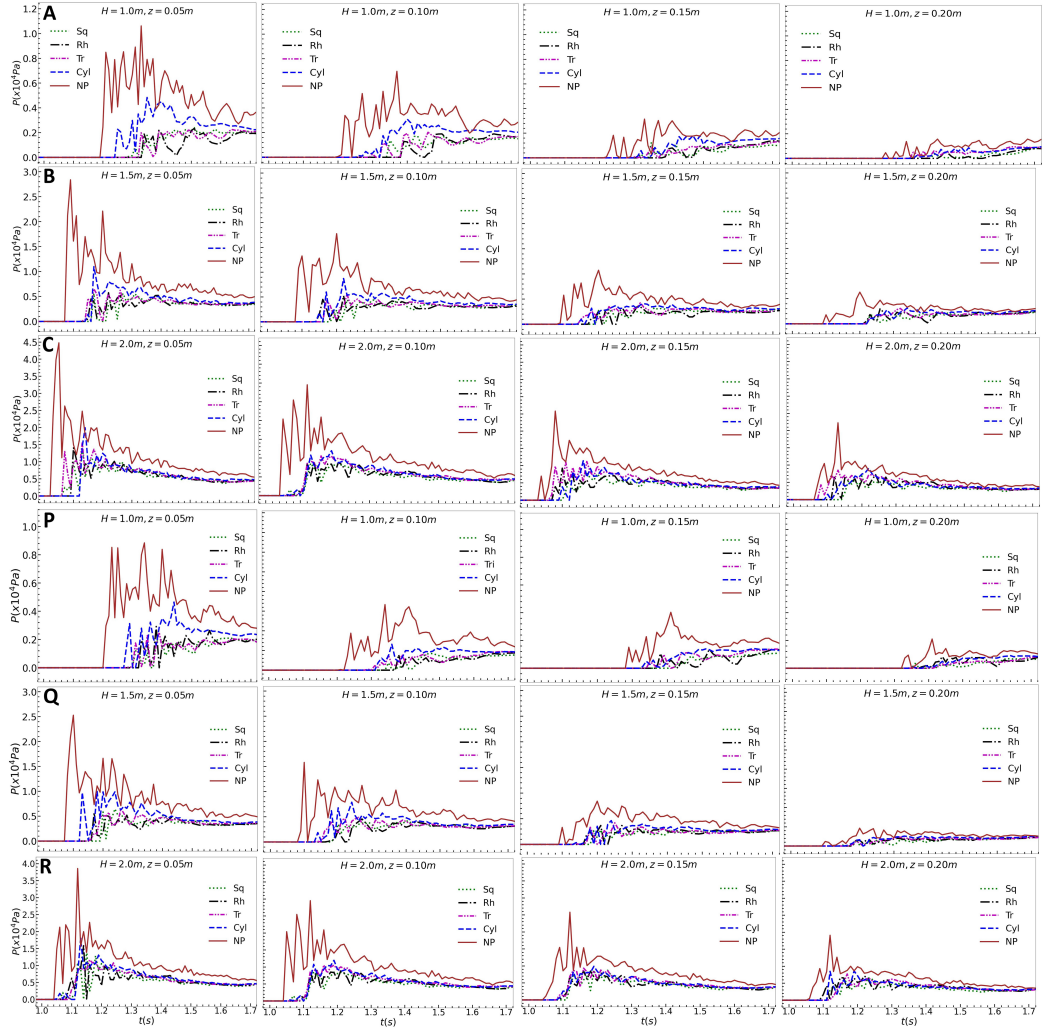}
	\caption{Pressure on the wall at $y=0.5$m for different pillar shapes for WBC (first to third rows) and PBC (fourth to sixth row).}
	\label{Fig:PresShape1}
\end{figure}
\noindent Whereas, in the case of triangles and rhombuses, the velocities are not reduced much as the flow faces a sharp edge, while inclined plane (with respect to the vertical plane parallel to the incoming flow direction) surfaces meet. 

%Fig14
Figure \ref{Fig:PresShape1} shows that, for all given dam heights, the cylindrical pillar has the least effective reduction in pressure on the wall, followed by the triangular, rhombus, and square pillars. A detailed analysis of the initial impact region $z = 0.1$m reveals that the square pillar exhibits the most significant reduction in pressure compared to the other pillars. 
The split flows passing by the square pillar are guided by the planes parallel to the side boundaries, giving rise to an almost rectangular shadow which cannot be easily filled. 
Hence, it offers a significant reduction in $P_{max}$ and protects the structure within the shadow for an extended period. For the triangle, two inclined planes direct the flows towards the side boundaries, which converge on the shadowed portion and merge, leading to an increase in pressure on the stopping wall. The cylindrical shape offers the highest pressure on the wall, as the round surface guides the split flows to the backside of the pillar, which contributes, along with the flows coming back from the side walls, to filling up the shadow.
In WBC, the pressure pattern follows the same trend as observed in PBC. However, the pressures with WBC are slightly higher.
%\textbf{The square pillar is the best one to obstruct the flow. The least pressure on the wall is provided by cylindrical pillar. The obtained results }
%
% Tab4
Table \ref{Tab:PeakPresShape} displays the values of $P_{max}^{PBC}$ and $P_{max}^{WBC}$ on the wall at various $z$ along $y=0.5$m for pillars of different shapes (cylinder, triangle, rhumbas, and square) with the same hydraulic diameter, $D=0.3$m placed at $a=1.0$m. 
We observe that $P_{max}^{WBC}$ is greater than the corresponding  $P_{max}^{PBC}$ for all cases shown here.
We also notice that both $P_{max}^{PBC}$ and $P_{max}^{WBC}$  at a particular $z$ increase with the increase in $H$ for all shapes. For a specific $H$ and shape, $P_{max}$ decreases with increasing $z$. At a particular $H$ and $z$, the pressure decreases to a maximum for a square shape and a minimum for a cylindrical shape, a trend that provides valuable insights into the behavior of different pillar shapes under varying conditions. 
Maximum percentage of reduction in $P_{max}$ in the presence of a pillar concerning $NP$ ($P_{max}^{NP}$) is about $70\%$ for the cases considered here.

% Tab5
Our calculations of $P_{max}$ at ($y=0.5$m, $z=1.0$m) on the front surface ($P_p$) of the pillar of various shapes, both for PBC and WBC, reveal interesting insights. We also present the maximum pressure on the wall ($P_w$) at the same ($y,z$) value. The results, tabulated in Table \ref {Tab5PpPw} considering the hydraulic diameter $D=0.3$m, $a=1.0$m, and $H=2.0$m, show that for both PBC and WBC, $P_p$ is smallest for the cylindrical shape because the water particles follow the circular path adhering to the geometry of the pillar and go to the back of the pillar (reduces the shadowed portion). For all other shapes, water particles are not allowed to go back of the pillar and hence this offer a larger $P_p$. The  largest pressure are observed on the square shape. The trend is reversed in the case of the values of $P_w$. This underscores the efficiency of the square pillar in protecting the structure behind it against the dam-break flow, along with the requirement of being strong enough to sustain the significant thrust caused by the flow. 

\begin{table}[H]
	\centering
	\caption{Maximum pressure value (in $Pa$) at   the front surface of the pillar ($P_p$) of $D=0.3$m  and on the wall ($P_w$) at $y=0.5$m,$z=0.1$m for different pillar shapes.}
	\label{Tab5PpPw}
	\begin{tabular}{|c|c|c|c|}
		\hline
		Shape & $P_p$ & $P_w$ & $P_w$/$P_p$ \\
		\hline
		Cyl & 16261.41 (\textcolor{blue}{17899.86
		}) & 11955.74 (\textcolor{blue}{15425.40}) & 0.74 (\textcolor{blue}{0.86}) \\
		\hline
		Tr  & 18439.76 (\textcolor{blue}{18128.58}) & 10576.75 (\textcolor{blue}{13520.41}) & 0.57 (\textcolor{blue}{0.75}) \\
		\hline
		Rh  & 19397.09 (\textcolor{blue}{19304.51}) & 8951.72  (\textcolor{blue}{12209.51}) & 0.46 (\textcolor{blue}{0.63}) \\
		\hline
		Sq  & 24534.36 (\textcolor{blue}{42617.62}) & 8611.20  (\textcolor{blue}{10977.81}) & 0.35 (\textcolor{blue}{0.49}) \\
		\hline
	\end{tabular}
\end{table}

\section{Conclusions}
We studied dam-break flow in the smoothed particle hydrodynamics framework using PBC for the domain side boundary which  corresponds to a wider flow, as a periodicity exists in the perpendicular direction of the flow, instead of usually employed rigid wall boundary plus slip/no-slip condition (WBC), which is appropriate for a narrow channel. 
We have assessed the change of the impact of the flow on the downstream structure (wall) due to the presence of an obstacle in front of it unlike the most of the reported works which deals the impact of the directly on the structure. 
We present the results of a systematic study on the effect of flow speed (related to $H$), the presence of obstacles in the flow pattern, and the pressure exerted on the wall behind the obstacle in the downstream region. We studied how the obstacle protects the structure behind it from the dam-break flow by changing the flow pattern and reducing the pressure on the wall for different sizes, locations with respect to the wall, and shapes of the obstructing pillar. We compared the results obtained using PBC and WBC for all the cases mentioned above. PBC exhibit a smaller impact on the structure compared to the WBC, where the particles are confined within the boundary.
For a tall dam ($H$), the potential energy is high, and hence the speed of the dam-break flow is large, resulting in a significant impact on the wall.
The pressure is large at the bottom of the wall and decreases with increasing height ($z$) in all cases. 
The presence of an obstacle in front of a structure on the downstream side splits the flow and creates a shadow behind the obstacle, which is observed to reduce the pressure on the structure by up to $70\%$ (for the cases considered here). 
The wider the obstacle, the greater is the reduction in pressure concerning the corresponding no pillar case (Np) than a thinner one.
Suppose an obstacle of the same size and shape is placed at a distant location from the wall. In that case, the flow split by the obstacle has the opportunity to return from the boundary and merge within the shadowed portion, resulting in a larger pressure at a point on the wall. Therefore, the closer the obstacle to a portion (object) to be protected, the less impact the portion will experience.
Regarding shape variation, it is observed that the least reduction in pressure occurs for a cylindrical pillar. Whereas,  the pressure at a point on the wall is reduced by a large amount for the square-shaped pillar. The impact at a point (with the same  $(y,z)$, as the point to be protected on the wall, but different $x$) on the front face of the pillar is largest in comparison to the other shapes considered here, implying that the strength of the pillar should be significantly more for this shape, in compared to the pillars of different shapes we considered.

It's important to note that we did not consider the turbulence term in our calculations. The inclusion of turbulence may alter the values of the calculated quantities. However, we believe that the overall trends and conclusions drawn from our study, particularly regarding the influence of obstacle size, shape, and location on pressure reduction, will remain similar even with the inclusion of turbulence.

We considered PBC which feel more appropriate for the systems considered here. Our aim was to provide systematic results on the impact on the wall due to a dam-break flow by altering various parameters of the obstacle placed in front of the wall. We estimate the parameters  of the obstacle best suited for maximum reduction of pressure on the structure. This study, with its practical implications, equips individuals with the knowledge to take preventive measures and protect structures more effectively and efficiently from disasters such as dam-break flows, flash floods, and tsunamis. 
 \bibliographystyle{elsarticle-num}
\bibliography{ref}% Produces the bibliography via BibTeX.

\end{document}